\documentclass[acmsmall,10pt]{acmart}

\acmJournal{PACMPL}
\acmVolume{1}
\acmNumber{CONF} 
\acmArticle{1}
\acmYear{2018}
\acmMonth{1}
\acmDOI{} 
\startPage{1}

\setcopyright{none}

\usepackage{booktabs}   
\usepackage{subcaption} 
\let\accentvec\vec

\usepackage{graphicx}
\usepackage{booktabs}   
\usepackage{subcaption} 
\usepackage{amsmath}
\usepackage{amssymb}
\usepackage{multicol}
\usepackage{amsthm}	
\usepackage{listings}
\usepackage{mathtools}
\usepackage{parcolumns}
\usepackage{scalerel}
\usepackage{color}
\usepackage{stackengine}
\usepackage{graphicx}
\usepackage{float}
\usepackage{tikz}
\usepackage{booktabs}
\usepackage{tcolorbox}
\usepackage{bbding}
\usepackage{wasysym}

\usepackage{multirow}
\usepackage[colorinlistoftodos]{todonotes}
\usepackage{lineno}
\usepackage{wrapfig}
\usepackage{centernot}
\usepackage{mathtools}
\usepackage{stmaryrd}
\usetikzlibrary{decorations.pathreplacing,calc}

\usepackage{pbox}

\usetikzlibrary{decorations.pathreplacing,calc}
\newcommand{\tikzmark}[1]{\tikz[overlay,remember picture] \node (#1) {};}

\newcommand*{\AddNote}[5]{%
	\begin{tikzpicture}[overlay, remember picture]
	\draw [decoration={brace,amplitude=0.3em},decorate, thick,black]
	($(#3)!([yshift=1.5ex]#1)!($(#3)-(0,1)$)$) --  
	($(#3)!(#2)!($(#3)-(0,1)$)$)
	node [align=center, text width=#4, pos=0.5, anchor=west] {#5};
	\end{tikzpicture}
}%

\newcommand*{\AddRec}[2]{%
	\begin{tikzpicture}[overlay, remember picture]
\draw [blue,thick,rounded corners] ($(#1)+(0,0.28cm)$) rectangle ($(#2)+(0,-0.1cm)$);
	\end{tikzpicture}
}%

\def\codec#1{\texttt{#1}}

\def\kname{$\mathbb{K}$}

\def\cd#1{\texttt{#1}}

\def\refrule#1{\hyperref[#1]{\textsc{#1}}}

\newcommand\RuleEnv[2]{
	\begin{center}
		\leavevmode
		\hbox{{\small
				\begin{tabular}{c}
					\multicolumn{1}{l}{\textsc{\textcolor{blue}{#1:}}}\\
					#2
		\end{tabular}}}
	\end{center}
}

\newcommand\RuleEnvShort[2]{
	\begin{center}
		\leavevmode
		\hbox{{\small
				\textsc{\textcolor{blue}{#1:}} 
				\begin{tabular}{c}
					#2 
		\end{tabular}}}
	\end{center}
}

\newcommand\ruleEnv[2]{
	\leavevmode
	\hbox{{\small
			\begin{tabular}{c}
				\multicolumn{1}{l}{\textsc{\textcolor{blue}{#1:}}}\\
				#2
	\end{tabular}}}
}

\newsavebox{\fmbox}

\newcommand\TwoColumn[2] {
	\begin{center}
		\begin{tabular}{cc}
			#1 & #2 
	\end{tabular}	\end{center}
}

\definecolor{mygreen}{rgb}{0,0.6,0}
\definecolor{mygray}{rgb}{0.5,0.5,0.5}
\definecolor{mymauve}{rgb}{0.58,0,0.82}

\def\rmem{\rightsquigarrow_m}

\makeatletter
\newcommand{\rmemm}[2][]{~\ext@arrow 0599{\Mapstofill@}{#1}{#2}_m}
\def\Mapstofill@{\arrowfill@{}\Relbar\Rightarrow}
\makeatother

\let\vec\accentvec

\newcommand\revision[1]{\textcolor{red}{#1}}

\def\cfg#1{\langle #1 \rangle}

\begin{document}

\title{An Executable Operational Semantics for Rust with the Formalization of Ownership and Borrowing}         


\author{Shuanglong Kan}
\orcid{nnnn-nnnn-nnnn-nnnn}             
\affiliation{
  \position{Position1}
  \department{Department of Computer Science}              
  \institution{TU Kaiserslautern}            
  \country{Germany}                    
}
\email{kanshuanglong@outlook.com}          

\author{Zhe Chen}
\orcid{nnnn-nnnn-nnnn-nnnn}             
\affiliation{
  \position{Position2a}
  \department{College of Compute Science and Technology}             
  \institution{NUAA}           
  \country{China}                   
}

\author{David San{\'a}n}
\affiliation{
  \institution{NTU}           
  \country{Singapore}                   
}

\author{Shang-wei Lin}
\affiliation{
	\institution{NTU}           
	\country{Singapore}                   
}

\author{Yang Liu}
\affiliation{
	\institution{NTU}           
	\country{Singapore}                   
}

\begin{abstract}

Rust is an emergent systems programming language highlighting memory safety by its Ownership and Borrowing System (OBS).
The existing formal semantics for Rust only covers limited subsets of the major language features of Rust.
Moreover, they formalize OBS as type systems at the language-level, which can only be used to conservatively analyze programs against the OBS invariants at compile-time. That is, they are not executable, and thus cannot be used for automated verification of runtime behavior.

In this paper, we propose RustSEM, a new executable operational semantics for Rust.
RustSEM covers a much larger subset of the major language features than existing semantics.
Moreover, RustSEM provides an operational semantics for OBS at the memory-level, which can be used to verify the runtime behavior of Rust programs against the OBS invariants.
We have implemented RustSEM in the executable semantics modeling tool K-Framework.
We have evaluated the semantics correctness of RustSEM wrt. the Rust compiler using around 700 tests.
In particular, we have proposed a new technique for testing semantic consistency to ensure the absence of semantic ambiguities on all possible execution selections. 
We have also evaluated the potential applications of RustSEM in automated runtime and formal verification for both functional and memory properties.
Experimental results show that RustSEM can enhance the memory safety mechanism of Rust, as it is more powerful than OBS in detecting memory errors.
	
\end{abstract}



\maketitle

\begin{acks}                            
  This material is based upon work supported by the
  \grantsponsor{GS100000001}{National Science
    Foundation}{http://dx.doi.org/10.13039/100000001} under Grant
  No.~\grantnum{GS100000001}{nnnnnnn} and Grant
  No.~\grantnum{GS100000001}{mmmmmmm}.  Any opinions, findings, and
  conclusions or recommendations expressed in this material are those
  of the author and do not necessarily reflect the views of the
  National Science Foundation.
\end{acks}

\section{Introduction}

Developing a formal semantics for a programming language could provide a mathematical foundation for the language. The semantics can be used as a reference model, and more importantly, used for proving language-level properties and constructing automated verification tools.

Rust~\cite{rusthome} is an emergent systems programming language aiming at providing memory safety guarantees with its Ownership and Borrowing System (OBS).
One of the most important guarantees maintained by the OBS invariants is the exclusive mutation capability for memory locations, which can avoid a variety of memory errors such as dangling pointers and data races.

Several formal semantics for Rust has been developed. Unfortunately, they only cover  limited subsets of the major language features of Rust, and most of them formalize either a variant or an outdated version of Rust.
For instance, Patina \cite{Eric2015} only captures the key features related to memory safety, e.g., unique pointers and references, in an early version of Rust.
RustBelt \cite{Ralf2018} formalizes a variant $\lambda_{Rust}$ of Rust in Coq, to prove in Coq that the type system of Rust can guarantee the memory and thread safe of $\lambda_{Rust}$ programs. 

Furthermore, most of the existing semantics for Rust is not executable, and thus cannot be used for automated verification of runtime behavior.
For instance, Patina \cite{Eric2015} and RustBelt \cite{Ralf2018} formalize OBS as type systems at the language-level, which can only be used to conservatively analyze programs against the OBS invariants at compile-time. But this may reject some execution traces satisfying the OBS invariants.
As a result, automated verification tools for Rust have to translate programs into the input languages of general-purpose verifiers, which may hurt expressiveness and make potential verification optimizations impossible.
For instance, Prusti \cite{AstrauskasMuellerPoliSummers19b} translates a subset of safe constructs into the input language of Viper \cite{mcai/0001SS16}.
To support direct verification of Rust programs, a new executable semantics should be developed for automated verification.

In this paper, we propose RustSEM, a new executable operational semantics for Rust. 
RustSEM consists of three levels:
\begin{enumerate}
	\item A sequential consistency memory model, which defines the memory layout of stacks and heaps, and a collection memory operations.
	More importantly, we formalize an \emph{operational semantics of OBS} in the memory model for dynamically checking the OBS invariants.
	Unlike the existing semantics of OBS, which is formalized as type systems at the language-level and bound to a specific language, our semantics is language-independent and thus can be reused for other languages since most languages have similar memory operations.
	Moreover, we also propose a high-level abstraction of OBS and prove the refinement relation between the abstraction and the memory model, ensuring the semantics correctness of the OBS refinement.
	\item An operational semantics for the Core Language (CL), an intermediate representation (IR) for reducing redundant semantics definitions.
	Recall that Rust also has an IR, namely Mid-level Intermediate Representation (MIR) \cite{MIR}. But the gap between Rust and CL is much smaller than MIR, facilitating the translation semantics from Rust to CL.
	\item A translation semantics from Rust to CL, which automatically translates
	Rust programs into the corresponding CL programs.
\end{enumerate} 
To execute a Rust program, RustSEM first translates the program into a CL program, and then executes the CL program wrt. the CL semantics, in which all memory accesses are carried out by invoking the interfaces of the memory model.

RustSEM distinguishes from the existing semantics in the following aspects.

Firstly, RustSEM directly formalizes Rust, instead of a variant, and covers a much larger subset of the major language features than existing semantics. For instance, RustSEM supports both safe and unsafe constructs, concurrency, dynamic OBS, closures, pattern matching and polymorphism.

Secondly, RustSEM provides an operational semantics for OBS at the memory-level, whilst other existing semantics formalizes OBS as type systems at the language-level.
The operational semantics of OBS is useful in two aspects.
(1) It can be used to verify the runtime behavior of Rust programs against the OBS invariants, i.e., reject the programs violating the OBS invariants. Compared with the type systems, it provides a more flexible OBS invariant checking that accepts a larger set of correct programs.
(2) It can be used to detect undefined behavior of the programs mixing safe and unsafe operations, by noting that unsafe constructs can escape from OBS and make programs prone to bugs \cite{Rustlibrarybug}.

Thirdly, RustSEM is an executable semantics, which means that RustSEM can execute a Rust program with respect to its semantics.
We have implemented RustSEM in the executable semantics modeling tool K-Framework (\kname) \cite{rosu-serbanuta-2010-jlap}.
\kname~is based on rewriting-logic and has a builtin parser that enables the semantics being defined on the abstract syntax tree. 
{\kname~has been successfully applied in formalizing the semantics of real-world programming languages, such as Java~\cite{bogdanas-rosu-2015-popl} and C~\cite{hathhorn-ellison-rosu-2015-pldi,ellison-rosu-2012-popl}}.
\kname's execution and verification engines enable testing RustSEM and constructing automated verifiers.

We have evaluated the semantics correctness of RustSEM wrt. the Rust compiler using around 700 tests for the three levels, {which mainly come from the Rust benchmarks \cite{benchmark}, the Rust libraries and the Rust text book \cite{rustbook}.}
Note that two semantics rules sometimes can be applied to the same construct at the same time, causing semantic ambiguities. Although one of them should be avoided, the execution engine of \kname~ always selects one rule to execute, leaving such ambiguities undetected.
Thus, we have proposed a new technique for testing semantic consistency to ensure the absence of semantic ambiguities on all possible execution selections, by exploiting the verification engine of \kname. 
Our new technique has discovered more than 36 semantic ambiguities.
{We also created tests for investigating what kinds of programs are rejected by the Rust compiler but accepted by RustSEM due to the compiler's conservation, and the reverse due to mixed safe and unsafe operations.}

We have also evaluated the potential applications of RustSEM in automated runtime and formal verification for both functional and memory properties.
Runtime verification with RustSEM is evaluated on 118 programs for detecting memory errors.
Formal verification with RustSEM is evaluated on a collection of benchmarks, including \codec{Vec\_Deque} in the Rust library implementing a ring buffer.
Experimental results show that RustSEM can enhance the memory safety mechanism of Rust, as it is more powerful than OBS in detecting memory errors.

In summary, we make the following contributions:
\begin{enumerate}
	\item We propose a new executable operational semantics for Rust that supports a \emph{larger} subset of the major language features, compared with all existing works.
	\item We propose a high-level abstraction of OBS and an operational semantics of OBS in the memory model. We also prove the refinement relation between them.
	\item We formalize the semantics of unsafe pointers and the semantics of the executions mixing both safe and unsafe pointers.
	\item We propose a novel testing technique based on \kname's verification engine to detect semantic ambiguities.
	\item We show that RustSEM can be applied to both runtime and formal verification against both functional and memory properties.  
\end{enumerate}
%


This paper is organized as follows:
Section \ref{sec:owner-borrow} recalls the OBS of Rust. 
Section \ref{sec:high-levelsemantics} presents a high-level abstraction of OBS.
Section \ref{sec:mem-model} defines the operational semantics of the memory model, while
Section \ref{sec:rustsemacs} presents the basic idea of the semantics of CL and the translation semantics for Rust.
Section \ref{sec:evaluation} evaluates the proposed semantics.
Section \ref{sec:relatedwork} compares related work.
Section \ref{sec:conclusion} concludes.

\section{Ownership and Borrowing System}
\label{sec:owner-borrow}

In this section, we recall the OBS of Rust and the related OBS invariants.

\subsection{Ownership} 

A variable can declare the unique \emph{ownership} of a memory block using a binding or an assignment. If the ownership is declared as mutable with the \cd{mut} keyword, then the owner can be used to both read and write the block, otherwise it is read-only.
A read-only (resp. mutable) owner is called a \emph{shared} (resp. \emph{mutable}) \emph{alias} of the block.
We denote by $x \rightarrow_o \cd{B}$ that variable \cd{x} is the owner of block \cd{B}, i.e., \cd{x} owns $\cd{B}$.
For instance, in Listing \ref{list:example-ownership}, the binding ``\cd{let mut v = vec![1,2]}" at Line \textcolor{blue}{1.1} first allocates a block $\cd{B}$ in the memory to store the vector \cd{[1,2]}, and then the owner \cd{v} obtains the ownership of $\cd{B}$, denoted by $\cd{v} \rightarrow_o \cd{B}$.

{\renewcommand\thelstnumber{%
		\ifnum\value{lstnumber}=1 1.1
		\else
		\ifnum \value{lstnumber}=2 1.2
		\fi
		\ifnum \value{lstnumber}=3 1.3
		\fi
		\ifnum \value{lstnumber}=4 1.4
		\fi
		\ifnum \value{lstnumber}=5 1.5
		\fi
		\ifnum \value{lstnumber}=6 1.6
		\fi
		\fi}
	
	\lstset{
		commentstyle=\color{red}, 
		numberstyle=\color{blue}\bfseries
	}

\begin{wrapfigure}{l}{0.45\textwidth} 
	\vspace{-0.4cm}
	{\small
		\quad\quad
		\begin{minipage}[t]{0.45\textwidth}
			\renewcommand{\ttdefault}{pcr}
			\begin{lstlisting}[caption=~\label{list:example-ownership},mathescape]
let mut v = vec![1,2];  $\tikzmark{own11}$
{
  let v1 = v;      $\tikzmark{own1}$ $\tikzmark{own12}$
  let t = v1[0];
  v1[1] = 3;
}   $\tikzmark{own2}$
			\end{lstlisting}
		\end{minipage}
\AddNote{own1}{own2}{own1}{0.55cm}{\cd{v1}}
\AddNote{own11}{own12}{own11}{0.55cm}{\cd{v}}}
\end{wrapfigure} 

An ownership can be \emph{moved} from one variable to another.
Moving an ownership from variable $x$ to another variable $y$
means that the ownership now belongs to $y$ and $x$ no longer owns it.
For instance, in Listing \ref{list:example-ownership}, the binding ``\cd{let v1 = v}'' at Line \textcolor{blue}{1.3} moves the ownership of the vector from \cd{v} to \cd{v1}, i.e., \cd{v1} becomes the new owner of the vector and \cd{v} can no longer be used to access it. Indeed, the vector is read and written through \cd{v1} at Lines \textcolor{blue}{1.4} and \textcolor{blue}{1.5}, respectively.

The \emph{lifetime} of an owner begins from the timestamp at which it obtains the ownership and ends at the timestamp at which it loses the ownership, e.g., when the ownership is moved or it goes out of the program scope (i.e., curly braces).
{Timestamps are a way to distinguish the execution order of program statements. In this section, we use line numbers as timestamps.}
The block is deallocated through its owner when the owner goes out of scope.
For instance, in Listing \ref{list:example-ownership}, the lifetime of owner \cd{v} begins at Line \textcolor{blue}{1.1} and ends at Line \textcolor{blue}{1.3}, whilst the lifetime of owner \cd{v1} begins at Line \textcolor{blue}{1.3} and ends at Line \textcolor{blue}{1.6}.
The vector is deallocated through \cd{v1} at Line \textcolor{blue}{1.6} as \cd{v1} goes out of scope.

\subsection{Borrowing and Reborrowing}

\emph{Borrowing} is a way to create \emph{references} to the owner of a memory block.
There are two kinds of references: \emph{shared} references (read-only, created by \cd{\&}) and \emph{mutable} references (readable and writable, created by \cd{\&mut}).
A shared (resp. mutable) reference is also called a \emph{shared} (resp. \emph{mutable}) \emph{alias} of the block.
A borrowing creating a shared (resp. mutable) reference is called a shared (resp. mutable) borrowing.
We denote by $x\rightarrow_{s} y$ (resp. $x\rightarrow_m y$) that $x$ is a shared (resp. mutable) reference to $y$, i.e., $x$ borrows $y$.
For instance, in Listing \ref{list:example-ref}, the borrowing ``\cd{let b1 = \&v1}" at Line \textcolor{blue}{2.3} creates a shared reference \cd{b1} to \cd{v1}, denoted by $\cd{b}1 \rightarrow_s \cd{v1}$, where \cd{b1} can only be used to read the block owned by \cd{v1}.
The borrowing ``\cd{let b2 = \&mut v2}" at Line \textcolor{blue}{2.4} creates a mutable reference, denoted by $\cd{b2} \rightarrow_m \cd{v2}$, where \cd{b2} can be used to both read and write the block owned by \cd{v2}.

	{\renewcommand\thelstnumber{%
		\ifnum\value{lstnumber}=1 2.1
		\fi
		\ifnum \value{lstnumber}=2 2.2
		\fi
		\ifnum \value{lstnumber}=3 2.3
		\fi
		\ifnum \value{lstnumber}=4 2.4
		\fi
		\ifnum \value{lstnumber}=5 2.5
		\fi
		\ifnum \value{lstnumber}=6 2.6
		\fi
		\ifnum \value{lstnumber}=7 2.7
		\fi
	}{\small
\quad
	\begin{minipage}[t]{0.45\textwidth}
		\renewcommand{\ttdefault}{pcr}
		\begin{lstlisting}[caption=~\label{list:example-ref},mathescape]
let v1 = vec![1,2];
let mut v2 = vec![1,2];               
let b1=&v1;     $\tikzmark{r1}$
let b2=&mut v2;          $\tikzmark{r2}$
let t1=(*b1)[0];    
let t2=(*b1)[1];    $\tikzmark{r1e}$
(*b2)[1] = 2;           $\tikzmark{r2e}$

		\end{lstlisting}
	\end{minipage}
	\AddNote{r1}{r1e}{r1}{1.5cm}{$\cd{b1}\rightarrow_s\cd{v1}$}
	\AddNote{r2}{r2e}{r2}{1.6cm}{$\cd{b2}\rightarrow_m\cd{v2}$}}
	\quad\quad
	{\renewcommand\thelstnumber{%
				\ifnum\value{lstnumber}=1 3.1
				\fi
				\ifnum \value{lstnumber}=2 3.2
				\fi
				\ifnum \value{lstnumber}=3 3.3
				\fi
				\ifnum \value{lstnumber}=4 3.4
				\fi
				\ifnum \value{lstnumber}=5 3.5
				\fi
			
				\ifnum \value{lstnumber}=8 3.6
				\fi
	}
	{\small
		\begin{minipage}[t]{0.43\textwidth}
			\renewcommand{\ttdefault}{pcr}
			\begin{lstlisting}[caption=~\label{list:example-ref4},mathescape]
let mut v = vec![1,2];
let b1=&mut v; $\tikzmark{l41}$
let b2=&(*b1);         $\tikzmark{l42}$
let z=(*b2)[0];
(*b1)[0] = 2;       $\tikzmark{l411}$
$\tikzmark{4tl}$/*b1 should be 
disabled by b2*/ $\tikzmark{4br}$
let t = (*b2)[1];           $\tikzmark{l421}$
			\end{lstlisting}
		\end{minipage}
		\AddNote{l41}{l411}{l41}{1.4cm}{$\cd{b1}\rightarrow_m\cd{v}$}
		\AddNote{l42}{l421}{l42}{1.5cm}{$\cd{b2}\rightarrow_s\cd{b1}$}
		\AddRec{4tl}{4br}}

\emph{Reborrowing} is a way to create references from another reference, instead of an owner. 
For instance, in Listing \ref{list:example-ref4}, 
\cd{b1} borrows the owner \cd{v} at Line \textcolor{blue}{3.2} and
\cd{b2} reborrows \cd{b1} with the referent $* \cd{b1}$ at Line \textcolor{blue}{3.3},
denoted by the link $\cd{b2}\rightarrow_s$  $ \cd{b1}\rightarrow_m \cd{v}$.
To access the first element of 
the vector, we can use \cd{v[0]}, \cd{(*b1)[0]} or \cd{(*b2)[0]}, where \cd{v}, \cd{*b1} and \cd{*b2} are three paths to access the vector.
Each path uses an alias as the entry to access the block.
For instance, the paths \cd{v}, \cd{*b1} and \cd{*b2} use the alias entries \cd{v}, \cd{b1} and \cd{b2}, respectively.

As borrowing and reborrowing are similar, we define unified relations for them.
\begin{definition} [Unified borrowing relation]
	Let $x$ and $y$ be two variables, where $y$ can be either an owner or a reference. 
	$x$ borrows or reborrows $y$ is denoted by ${x}\rightarrow_b{y}$. 
	The borrowing relation $\rightarrow_b$ contains two sub-relations: \emph{shared} borrowing $\rightarrow_s$ and \emph{mutable} borrowing $\rightarrow_m$, i.e., $ x \rightarrow_b y \Leftrightarrow (x\rightarrow_s y) \lor (x\rightarrow_m y)$.
\end{definition}

The \emph{lifetime} of a reference begins from the timestamp of its creation and ends at the last timestamp at which it is used (read, written, or borrowed), according to the definition of Non-Lexical Lifetimes (NLL) \cite{NLL2018}. 
Unlike the lifetime of owner, which can be easily decided by the program scope, the lifetime of reference is decided by the last use of it.
For instance, in Listing \ref{list:example-ref}, the lifetime of \cd{b1} is from Line \textcolor{blue}{2.3} to \textcolor{blue}{2.6}, while the lifetime of $\cd{b2}$ is from Line \textcolor{blue}{2.4} to \textcolor{blue}{2.7}.

\subsection{The OBS Invariants}

Because a memory block can be accessed by multiple aliases, OBS should fulfill the following OBS guarantees in order to avoid memory errors. 
\begin{itemize}
	\item Each alias only accesses (reads or writes) one valid block. For owners, they should own the block. For references, their owners should be in their lifetimes to ensure validity.
	\item At any time in an execution, each block can be accessed by either multiple shared aliases (but no mutable alias) or exclusively	accessed by a unique mutable alias.
\end{itemize}

In order to meet these two guarantees, OBS maintains the following invariants.
We will prove that these invariants exactly ensure the guarantees in Section \ref{sec:high-levelsemantics}.

\begin{definition} [The OBS invariants]
	\label{def-rule-unified}
	The following invariants must be satisfied:
	\begin{enumerate}
		\item \emph{Unique owner invariant}: Each block has a unique owner.
		\item \emph{Lifetime inclusion invariant}: If ${x}\rightarrow_b{y}$ then the lifetime of $x$ should always be within the lifetime of $y$ in order to avoid dangling pointers.
		\item \emph{Lifetime disjoint invariant}: There are \emph{no} two references to the same referent such that their lifetimes intersect and one of them is a mutable reference.
		\item \emph{Writing permission invariant}: If ${x}\rightarrow_s {y}$ then the writing permission of \cd{y} should be disabled until the end of $x$'s lifetime.
		\item \emph{Reading and writing permission invariant}: If ${x}\rightarrow_m {y}$ then both the reading and writing permissions of \cd{y} should be disabled until the end of $x$'s lifetime.
	\end{enumerate}
\end{definition}

We now illustrate the above invariants by examples.
The unique owner invariant is obvious. The move operation preserves the invariant.
Listing \ref{list:example-ref5} violates Invariant (2).
The lifetime of owner \cd{v} is from Line \textcolor{blue}{4.3} to \textcolor{blue}{4.6}
and the lifetime of reference $\cd{b}$ is from Line \textcolor{blue}{4.4} to \textcolor{blue}{4.7}.
Therefore the reference's lifetime is not within the owner's lifetime.

		{\renewcommand\thelstnumber{%
					\ifnum\value{lstnumber}=1 4.1
					\fi
					\ifnum \value{lstnumber}=2 4.2
					\fi
					\ifnum \value{lstnumber}=3 4.3
					\fi
					\ifnum \value{lstnumber}=4 4.4
					\fi
					\ifnum \value{lstnumber}=5 4.5
					\fi
					\ifnum \value{lstnumber}=6 4.6
					\fi
					\ifnum \value{lstnumber}=7 4.7
					\fi
					
				}
	\quad{\small
				\begin{minipage}[t]{0.45\textwidth}
					\renewcommand{\ttdefault}{pcr}
					\begin{lstlisting}[caption=~\label{list:example-ref5},mathescape]
let mut b;
{ 
  let mut v = vec![1,2];   $\tikzmark{l61}$
  b = & mut v;    $\tikzmark{l51}$
  let t = (*b)[1];
}       $\tikzmark{l62}$
(*b)[1] = 2;           $\tikzmark{l52}$
					\end{lstlisting}
				\end{minipage}
				\AddNote{l61}{l62}{l61}{0.4cm}{\cd{v}}
				\AddNote{l51}{l52}{l51}{1.1cm}{$\cd{b}\rightarrow_s\cd{v}$}}
	{\renewcommand\thelstnumber{%
		\ifnum \value{lstnumber}=2 5.1
		\fi
		\ifnum \value{lstnumber}=3 5.2
		\fi
		\ifnum \value{lstnumber}=4 5.3
		\fi
		\ifnum \value{lstnumber}=5 5.4
		\fi
		\ifnum \value{lstnumber}=6 5.5
		\fi
		\ifnum \value{lstnumber}=7 5.6
		\fi
		\ifnum \value{lstnumber}=8 5.6
		\fi
	}
	{\small
		\quad
		\begin{minipage}[t]{0.45\textwidth}
			\renewcommand{\ttdefault}{pcr}
			\begin{lstlisting}[caption=~\label{list:example-ref1},mathescape]
struct PT{x:i32,y:i32}
let mut v=PT{x:1,y:2};
let b1=&v;       $\tikzmark{b1-1}$
let b2=&mut(v.x);       $\tikzmark{b2-1}$
let t=(*b1).x;  $\tikzmark{b1-2}$
(*b2) = 2;               $\tikzmark{b2-2}$
			\end{lstlisting}
		\end{minipage}
		\AddNote{b1-1}{b1-2}{b1-1}{1.2cm}{$\cd{b1}\rightarrow_s$\cd{v}}
		\AddNote{b2-1}{b2-2}{b2-1}{1.5cm}{$\cd{b2}\rightarrow_m$\cd{v.x}}}

Listing \ref{list:example-ref1} violates Invariant (3). 
It is a more interesting example as
\cd{b2} borrows the field \cd{x} of \cd{v} instead of \cd{v}.
Rust also enables fine-grained borrowings for \emph{struct} types.
Let \cd{ST} be a \emph{struct} type with fields $d_1,\ldots,d_n$ and 
\cd{x} be the owner of a block of the type \cd{ST}.
We can create borrowings for both \cd{x} (e.g. $\cd{y}\rightarrow_b \cd{x}$) and the fields of \cd{x}  (e.g. $\cd{y}\rightarrow_b \cd{x}.d_i$) .
The relation between them is:
$
\cd{y}\rightarrow_b \cd{x} \Rightarrow y\rightarrow_b \cd{x}.d_1\land \ldots \land
y\rightarrow_b \cd{x}.d_n.
$
This means that if \cd{x} is borrowed then all its fields are borrowed.
It also means that  an alias can not only be a variable but also be fields of variables.
Moreover, a block could be divided into ``sub-blocks'' corresponding fields.
The lifetime of  $\cd{b1}\rightarrow_b \cd{v}$ is from Line \textcolor{blue}{5.2} to \textcolor{blue}{5.4}.
We can also infer that $\cd{b1}\rightarrow_s \cd{v.x}$ from $\cd{b1}\rightarrow_s \cd{v}$.
The lifetime of  $\cd{b2}\rightarrow_m \cd{v.x}$ is from Line \textcolor{blue}{5.3} to \textcolor{blue}{5.5}. 
The intersection of the lifetimes of $\cd{b1}\rightarrow_s \cd{v.x}$ and $\cd{b2}\rightarrow_m \cd{v.x}$ is not empty and one of them is a mutable reference. 
Listing \ref{list:example-ref4} violates Invariant (4).
The lifetime of shared reference \cd{b2} is from Line \textcolor{blue}{3.3} to  \textcolor{blue}{3.6}.
Since \cd{b2} reborrows \cd{b1}, \cd{b1}'s writing permission should be disabled within the lifetime of $\cd{b2}$.
Thus, the writing through \cd{b1} at Line  \textcolor{blue}{3.5} is illegal.

Both owners and references are safe pointers. But Rust also supports raw pointers, which are unsafe pointers that can escape from OBS checking.
We formalize raw pointers in the memory model in Section \ref{sec:mem-model}.
Besides raw pointers, Rust also has other unsafe features including unsafe scopes and unsafe functions, which are statement scopes and functions wherein raw pointers are read or written.
We model these unsafe features in CL and the translation semantics in Section \ref{sec:rustsemacs}.

\section{High-level Abstraction of OBS }
\label{sec:high-levelsemantics}
{
	This section presents a high-level formalization of OBS by graphs, in which we only focus on the ownership and borrowing relations among aliases and blocks.
	At the end of this section, we prove that the OBS invariants exactly ensure the OBS guarantees.}

{
	\begin{definition}[Lifetimes]
		Let $\emph{Tim}$ be an infinitely countable set of timestamps.
		$(\emph{Tim},\leq)$ is a totally ordered set.
		A strict order $<$ over $\emph{Tim}$ is defined as $t< t'$ iff $t\leq t'\land t\neq t'$.
		The function $\emph{Su}:\emph{Tim}\rightarrow\emph{Tim}$ is the successor function of timestamps,
		such that $\emph{Su}(t)=t'$ iff $t<t'\land\nexists t''. t< t'' < t'$.
		{A lifetime $t_1\sim t_2~(\text{where }t_1,t_2\in \emph{Tim})$, defined as $\{t\mid t_1\leq t\leq t_2\}$, is a subset of $\emph{Tim}$}.
\end{definition}}

{
		Let $LT$ denotes the set of all lifetimes.
	Because a lifetime is a set of timestamps, we can use set operations on lifetimes.
}

Blocks, aliases, and their relations constitute an OBS graph, defined as follows:
\begin{definition}[OBS graph of a block]
	An OBS graph is defined as $G=(B,V,E,\mathcal{F})$, where   
	\begin{itemize}
		\item $B$ is a memory block,
		\item $V$ is a set of aliases of $B$, e.g., owners and references,
		\item $E: V \rightarrow \{o,s,m\} \times (\{B\} \cup V)$ is a set of edges, i.e., a total mapping from aliases to the product of $\{o,m,s\}$ and $\{B\} \cup V$.
		$E(a)=(o,B)$ iff $a\rightarrow_o B$, i.e. $a$ owns $B$.
		$E(a)=(s,a')$ iff $a\rightarrow_s a'$, i.e., $a$ is a shared reference to $a'$.
		$E(a)=(m,a')$ iff $a\rightarrow_m a'$, i.e., $a$ is a mutable reference to $a'$.
		\item $\mathcal{F}: E\rightarrow LT$ is the lifetimes of edges.
	\end{itemize}	
\end{definition}

Note that $E$ is a total mapping as each alias holds exactly one value. In the sequel, $E(a)=(*,a')$ and $(a\rightarrow _* a')\in E$ are used interchangeably, where $* \in \{ o, s, m\}$.
{We also inherit the notation $\rightarrow_b$ in Section \ref{sec:owner-borrow}
	to denote either $\rightarrow_{m}$ or $\rightarrow_{s}$.}
Fig. \ref{fig:example-obg} is an example of OBS graphs.
On the left is a Rust program and right
is its two OBS graphs, in which one is for \cd{B} and 
another one is for \cd{B}.0.
 The symbol $\cd{B}$ denotes
 the block that stores the vector \cd{PT\{x:1,y:2\}} and
 $\cd{B}.0$ is the offset storing the value of the field \cd{x} and is a sub-block of \cd{B}.
 
We first explain the OBS graph for \cd{B}.
In the graph, an edge is labeled by the mutability of the borrowing and its lifetime.
For instance, the edge from \cd{b2} to \cd{b1} is labeled by $s$[3-4], it denotes
$\cd{b2}\rightarrow_s \cd{b1}$ and $\mathcal{F}(\cd{b2}\rightarrow_s \cd{b1})=\{3,4\}$. The notation [3-4] denotes the range from 3 to 4.
The second OBS graph of \cd{B}.0 is for the field \cd{x}, since \cd{b3} directly borrows \cd{v.x}.
The dashed arrow in the OBS graph is inferred from the borrowing $\cd{b}_1\rightarrow_m \cd{v}$.

To formalize the OBS invariants (1), (2) and (3), we define well-formedness for OBS graphs.

{\renewcommand\thelstnumber{%
		\ifnum \value{lstnumber}=2 1
		\fi
		\ifnum \value{lstnumber}=3 2
		\fi
		\ifnum \value{lstnumber}=4 3
		\fi
		\ifnum \value{lstnumber}=5 4
		\fi
		\ifnum \value{lstnumber}=6 5
		\fi
		\ifnum \value{lstnumber}=7 6
		\fi
		\ifnum \value{lstnumber}=8 7
		\fi
	}
\begin{figure}
	\centering
	{\small
		\begin{minipage}[t]{0.9\textwidth}
			\renewcommand{\ttdefault}{pcr}
			\begin{lstlisting}[mathescape]
struct PT {x:i32,y:i32}                $\tikzmark{obg}$
let mut v = PT{x:1,y:2};                 
let b1 = & mut v;         $\tikzmark{l41}$                
let b2 = & (* b1);$\tikzmark{l42}$                      
let z = (*b2).x;$\tikzmark{l421}$
(*b1).y = 2;             $\tikzmark{l411}$
let b3 = &mut (v.x); $\tikzmark{4tl}$   
(*b3) = 3;           $\tikzmark{4br}$

			\end{lstlisting}
		\end{minipage}
		
		\AddNote{l41}{l411}{l41}{1.4cm}{$\cd{b1}\rightarrow_m\cd{v}$}
		\AddNote{l42}{l421}{l42}{1.4cm}{{$\cd{b2}\rightarrow_s\cd{b}_1$}}
		\AddNote{4tl}{4br}{4tl}{1.6cm}{$\cd{b3}\rightarrow_m\cd{v.x}$}
		
		\begin{tikzpicture}[overlay, remember picture]
		\draw ($(obg)$) circle (0.25cm);
		\draw ($(obg)+(1.5cm,0)$) circle (0.25cm);
		\draw ($(obg)+(3cm,0)$) circle (0.25cm);
		\draw ($(obg)+(4.5cm,0)$) circle (0.25cm);
		\draw[thick,->] ($(obg)+(0.25cm,0)$) -- ($(obg)+(1.25cm,0)$) ;
		\draw[thick,->] ($(obg)+(1.75cm,0)$) --($(obg)+(2.75cm,0cm)$) ;
		\draw[thick,->] ($(obg)+(3.25cm,0)$) --($(obg)+(4.25cm,0cm)$) ;
		
		\node at ($(obg)$) {\cd{b2}};
		\node at ($(obg)+(1.5cm,0)$) {\cd{b1}};
		\node at ($(obg)+(3cm,0cm)$) {\cd{v}};
		\node at ($(obg)+(0.75cm,0.22cm)$) {{$s$[3-4]}};
		\node at ($(obg)+(2.2cm,0.22cm)$) {{$m$[2-5]}};
		\node at ($(obg)+(4.5cm,0cm)$) {\cd{B}};
		\node at ($(obg)+(3.75cm,0.22cm)$) {{$o$[1-7]}};
		
		\draw ($(obg)+(0,-1.5cm)$) circle (0.25cm);
		\draw ($(obg)+(1.5cm,-1.5cm)$) circle (0.25cm);
		\draw ($(obg)+(1.5cm,-2.5cm)$) circle (0.25cm);
		\draw ($(obg)+(3cm,-2cm)$) circle (0.25cm);
		\draw ($(obg)+(4.5cm,-2cm)$) circle (0.25cm);
		\draw[thick,->] ($(obg)+(0.25cm,-1.5cm)$) -- ($(obg)+(1.25cm,-1.5cm)$) ;
		\draw[dashed, thick,->] ($(obg)+(1.75cm,-1.5cm)$) --($(obg)+(2.75cm,-1.95cm)$) ;
		\draw[thick,->] ($(obg)+(1.75cm,-2.5cm)$) --($(obg)+(2.75cm,-2.05cm)$) ;
		\draw[thick,->] ($(obg)+(3.25cm,-2cm)$) --($(obg)+(4.25cm,-2cm)$) ;
		
		\node at ($(obg)+(0,-1.5cm)$) {\cd{b2}};
		\node at ($(obg)+(1.5cm,-1.5cm)$) {\cd{b1}};
		\node at ($(obg)+(1.5cm,-2.5cm)$) {\cd{b3}};
		\node at ($(obg)+(3cm,-2cm)$) {\cd{v.x}};
		\node at ($(obg)+(0.75cm,-1.3cm)$) {{$s$[3-4]}};
		\node at ($(obg)+(2.45cm,-1.4cm)$) {{$m$[2-5]}};
		\node at ($(obg)+(2.45cm,-2.6cm)$) {{$m$[6-7]}};
		\node at ($(obg)+(4.5cm,-2cm)$) {\cd{B}.0};
		\node at ($(obg)+(3.75cm,-1.7cm)$) {{$o$[1-7]}};
%
		
		\end{tikzpicture}}
	\vspace{-0.4cm}
	\caption{{An example of OBG graphs}}
	\label{fig:example-obg}
\end{figure}

	{\renewcommand\thelstnumber{%
		\ifnum\value{lstnumber}=1 1
		\else \arabic{lstnumber}
		\fi
	}

\begin{definition}[Well-formed OBS graphs]
	\label{def-well-formed-obs}
	Let $G=(V,B,E,\mathcal{F})$ be an OBS graph. It is well-formed if and only if it satisfies
	\begin{enumerate}
			\item There is a unique $a\in V$ such that $E(a)=(o,B)$.
		\item $\forall a, a', a'' \in V\cup\{B\}$. $E(a)=(*,a') \wedge E(a')=(*,a'')$ $\implies$ $\mathcal{F}(a\rightarrow_* a')\subset\mathcal{F}(a'\rightarrow_* a'')$.
		\item $\forall a, a', a'' \in V$. $E(a')=(b,a) \wedge E(a'')=(m,a)$ $\implies$ $\mathcal{F}(a'\rightarrow_b a) \cap \mathcal{F}(a''\rightarrow_m a)=\emptyset$.
	\end{enumerate}
\end{definition}
{

In Fig. \ref{fig:example-obg}, the two OBS graphs are well-formed.
For instance, the block \cd{B}.0 is uniquely pointed by \cd{v.x}.
The lifetime of $\cd{b2}\rightarrow_s \cd{b1}$ is within
the lifetime of $\cd{b1}\rightarrow_m \cd{v.x}$.
The lifetimes of $\cd{b}_1\rightarrow_m\cd{v.x}$ and
{
	$\cd{b}_3\rightarrow_m\cd{v.x}$ do not intersect.}
One important property for the  well-formed OBS graphs is acyclic.} The proof is in Appendix \ref{sec-acyclic}.

\begin{lemma}
	\label{lemma-acyclic}
	A well-formed OBS graph is acyclic.
\end{lemma}

To formalize the OBS invariants (4) and (5), we define reading and writing permission functions. An alias is permitted to read or write	at a specific timestamp if the reading or writing permission function returns true, respectively.
\begin{definition}[Permission functions]
	\label{def-permissionfunction}
Let $G=(B,V,E,\mathcal{F})$ be an OBS graph.
The reading permission function
$R_G:V\times \emph{Tim}\rightarrow \mathbb{B}$  is defined as 
	\[
	R_G(a,t) = \left\{\begin{array}{ll}
	\emph{true} & \emph{if}~~(\exists a'.  {a\rightarrow_* a'\in E}\land t\in \mathcal{F}(a\rightarrow_* a'))\land (\nexists a''. t\in\mathcal{F}(a''\rightarrow_m a))\\
	\emph{false} &  \emph{otherwise}
	\end{array} \right.
	\]
	where the condition $t\in \mathcal{F}(a\rightarrow_* a')$ denotes that $t$ is in the lifetime of the relation.
	Note that the condition $\nexists a''. t\in\mathcal{F}(a''\rightarrow_m a)$ ensures that $a$'s reading permission is not disabled at $t$.
	
	The writing permission function $W_G:V\times \emph{Tim}\rightarrow \mathbb{B}$ is defined as
	\[
	W_G(a,t) = \left\{\begin{array}{ll}
	\emph{true} & \emph{if}~{(\exists a'.  a\rightarrow_{\{o,m\}}a'\in E}\land t\in {\mathcal{F}(a\rightarrow_{\{o,m\}} a'))}\land {(\nexists a''. t\in\mathcal{F}(a''\rightarrow_b a))}\\
	\emph{false} &  \emph{otherwise}
	\end{array} \right.
	\]
	where $a\rightarrow_{\{o,m\}}a'$ denotes $a\rightarrow_o a'$ or $a\rightarrow_m a'$.
	
\end{definition}

{
In Fig. \ref{fig:example-obg},
at timestamp 3, 
$W_G(\cd{b1},3)=\emph{false}$ as there exists
$\cd{b2}$ such that $3\in\mathcal{F}(\cd{b2}\rightarrow_s\cd{b1})$.
At timestamp 7, $R_G(\cd{b3},7)=\emph{true}$ as $7\in\mathcal{F}(\cd{b3}\rightarrow_m\cd{v.x})$
and $\nexists a'. a'\rightarrow_b \cd{b3}\in E$.}

In the following theorem, we show that 
a well-formed OBS graph satisfies  exclusive mutation guarantee with respect to the permission functions.
The proof is in Appendix \ref{app-exclusive}.
\begin{theorem}\label{theorem-unifimodel}
	\label{theorem-exclusive}
	Let $G=(V,B,E,\mathcal{F})$ be an OBS graph and $t$ be a timestamp. If {$G$} is well-formed then we have either
	(1) $\forall a\in V, W_G(a,t)={\text{false}}$ or
	(2) $\exists! a\in V, W_G(a,t)=\text{true}\land (\forall a'. a\neq a'\Rightarrow R_G(a',t)=$ $W_G(a',t)=\text{false})$.
	The notation $\exists!$ denotes unique existential quantification.
\end{theorem}

Theorem \ref{theorem-exclusive} shows that for a block or a sub-block, 
if all readings or writings by the aliases in its OBG graph are executed at the timestamps
that the corresponding permission functions return true then
the readings and writings respect exclusive mutation guarantee.

{Compared with the OBS of Rust compiler, our formalization relaxes some restrictions.
For instance, in Rust compiler, an alias $a$ could write a block owned by $a'$ if all the borrowings in the path from $a$ to $a'$
	in the OBS graph are all mutable borrowings and $a'$ is declared as mutable.
In the high-level abstraction, we have no such restriction.
But in the memory model (Section \ref{sec:mem-model}) of RustSEM, we still respect this restriction.}


{
In Rust, there is another definition of the lifetimes for references, called two-phase borrowing \cite{twophase-borrowing},
in which the lifetime of a mutable reference is from the timestamp that we start to use it instead of the timestamp that it is created.
Our high-level abstraction is also compatible with it.
We only need to reconsider the lifetimes of references,
For instance, in Fig. \ref{fig:example-obg} $\mathcal{F}(\cd{b3}\rightarrow_m \cd{v.x})$ is $\{7\}$ now.
Definition \ref{def-well-formed-obs} and \ref{def-permissionfunction} keep unchanged and Theorem \ref{theorem-exclusive} is still true.
}




\section{Memory Model}
\label{sec:mem-model}

In this section, we introduce our memory model, which is the core of RustSEM that formalizes OBS.
{
The memory model supports the dynamic checking of OBS invariants.
It can be viewed as an implementation or a refinement of
the high-level abstraction of OBS.
Moreover, it also supports sequential consistency checking for concurrent access.
 }

We first illustrate our basic idea, especially the \emph{dynamic lifetime extension}, with a motivation example in Fig. \ref{fig:motivationexample}.
There are 7 lines of code in the figure.
{
Line 1 creates a block $B$ for \cd{vec![1,2]}, whose address is $\cd{b}$.
The symbol $\cd{v}\mapsto \cd{own}(\cd{b})$ means that
the variable $\cd{v}$ starts to own the block \cd{b}.
It denotes $\cd{v}\rightarrow_o B$.
The value $\cd{own}(\cd{b})$ indicates that the ownership of $B$ belongs to \cd{v}.}
Line 2 creates a borrowing relation $\cd{b1}\rightarrow_m \cd{v}$.
We introduce $\cd{mut}(2\sim 2, \cd{v})$ as a value to denote a \emph{reference value} to $\cd{v}$ in the memory model,
{where $2\sim 2$ is a timestamp span from Line 2 to 2 in which \cd{b1} is used}, since we only scan the code to
Line 2 until now.
When we scan Line 3, \cd{b1} is used to write the vector,
the {timestamp span} of the reference \cd{b1} should be extended since
it is used here. The new {span} is $2\sim 3$.
This treatment for lifetimes is called {the} \emph{dynamic lifetime extension}.
Line 4 writes the vector by its owner \cd{v}.
Note that $\cd{b1}$ does not disable \cd{v} here since 
its lifetime is not extended to Line 4.
Line 5 creates a relation $\cd{b2}\rightarrow_{s}\cd{v}$.
We introduce the value $\cd{shr}(5\sim5,\cd{v})$ to denote
a shared reference value to \cd{v}, whose lifetime is from Line 5 to 5 at the moment.
Line 6 writes the vector by its owner, this is allowed at the moment
since the existing two references cannot disabled it until now.
But at Line 7, since \cd{b2} is used {to read},
the lifetime of the reference \cd{b2}
should be extended to 7.
Now, an error arises, i.e, the writing at Line 6 is within the lifetime
of the shared reference, thus it is illegal.

\begin{wrapfigure}{r}{0.63\textwidth}
	\includegraphics[scale=0.5]{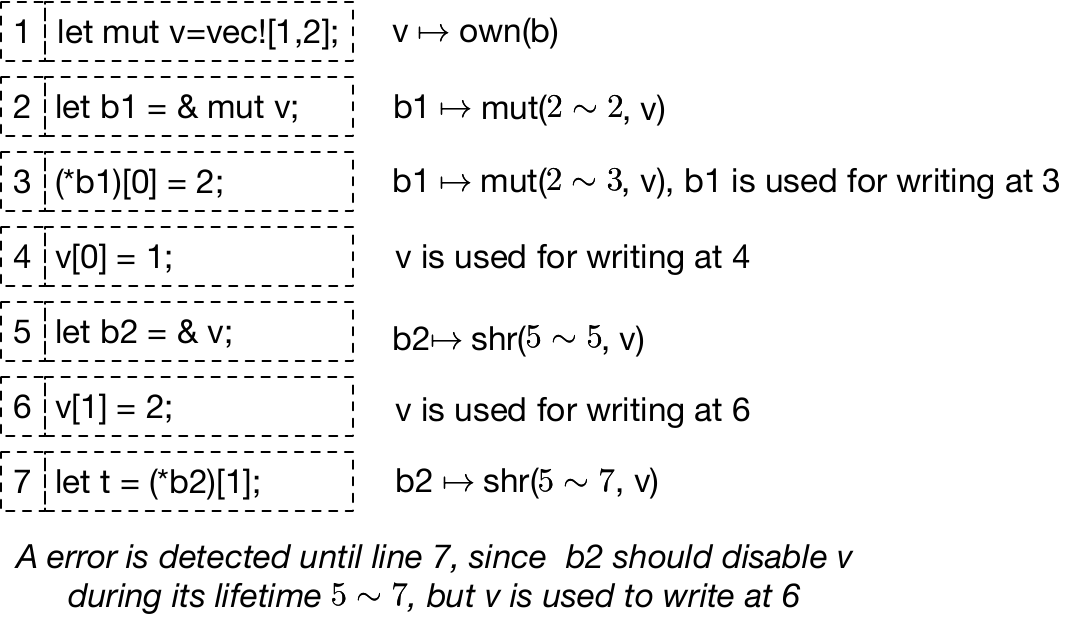}
	\caption{The motivation example}
	\label{fig:motivationexample}
\end{wrapfigure}

From the motivation example, in order to support dynamic OBS invariant checking without looking
backward, we select to store lifetimes for references. Moreover, we also
need to record at which timestamps an alias
is {used to read or write.}
For instance, in Fig. \ref{fig:motivationexample} the owner \cd{v} is recorded to be {used to write}
at Line 4 and 6.
Here we use lines to denote {timestamps} for illustration.
In the memory model, {timestamps} are generated by semantics rules.
Fig. \ref{fig:constructs-memorymodel} illustrates
the grammar of the memory model.
We detail them in the following subsections.

\begin{figure}
	\centering
	{\small
		\begin{minipage}{0.9\textwidth}
			\begin{tcolorbox}
				\[
				\begin{array}{lrcl}
				\text{Natural number} & \cd{n},\cd{m} \in \mathbb{N}  & & \\
				\text{Types} & \cd{t} \in \mathcal{T} & & \\
				\text{Time stamps} & \cd{ts} \in \textit{Tim} &  & \\
				\text{Local Lifetimes} & lt   \in LT &  \text{::=} & \cd{ts}_1\sim\cd{ts}_2\\
				\text{{Stack locations}} & \cd{s}\in L_s & & (L_s,\leq_s)\text{ is a }\textit{partially ordered set}\\
				\text{{Block locations}} & \cd{b}\in L_b &  &(L_b,\leq_b) \text{ is a }\textit{totally ordered set}\text{ and }L_s\cap L_b=\emptyset\\
				\text{Heap locations} & (\cd{b},\cd{n}) \in L_h  & &L_h\triangleq L_b\times\mathbb{N}\\
				\text{Mem locations} & l\in L_m  & ::= & \cd{s}\mid  (\cd{b},\cd{n}) \\
				\text{Paths} & p & ::= & \cd{s} \mid \cd{*} p \mid p\cd{.}\cd{n}\\
				\text{Alias} & \cd{a}\in \mathcal{A} & ::= &\cd{s}\mid \cd{a}.\cd{n} \\
				\text{Pointer Values} & pv & ::= & \cd{own}(\cd{b}) \mid \cd{shr}({lt}, p) \mid \cd{shr}(p)\\
				& & \mid & \cd{mut}({lt},p)\mid \cd{mut}(p)\mid \cd{raw}(l)\\
				\text{Primitive Values} & sv\in SV & ::= & i\in \cd{Int}\mid str\in\cd{Str}\mid c\in \cd{Char}
				\\
				&&\mid & f\in\cd{Float}\mid bv\in\cd{Bool} \\
				\text{Values} & v\in\cd{Val} & ::= & \bot\mid  pv \mid sv \\\\
				\hline
				\\
				\text{Mem Operations} &~~mop & ::=  & \cd{alloc(n,t)}\mid  \cd{free(b)}\\
				& & \mid &\cd{lv}(p)\mid\cd{read}(p)\mid \cd{write}(p,v) \\
				& & \mid & 
				\cd{rawRead}(l)\mid \cd{rawRead}'(l) \\
				& & \mid &  \cd{rawWrite}(l,v) \mid \cd{rawWrite}'(l,v) \\
				& & \mid & \cd{aRead}(l) \mid \cd{aWrite}(l,v)\\
				\end{array}
				\]
			\end{tcolorbox}
	\end{minipage}}
	\caption{The grammar of the memory model}
	\label{fig:constructs-memorymodel}
	\vspace{-0.5cm}
\end{figure}

\subsection{Memory Configuration}
\label{sub-memoryconfiguration}
A memory configuration is defined as a 4-tuple $mem=(S,H,\mathcal{P}, ms)$,
where $S$ is for stacks, $H$ is for a heap, $\mathcal{P}$ stores the latest timestamps at which an alias
is {used to read or write} as we explained in the motivation example, $ms$ is used for sequential consistency
checking of concurrent access. We elaborate them as follows.

The stacks are modeled as a finite partial map $S: L_s\xrightharpoonup{\emph{fin}} (\cd{Val}\times \mathcal{T})$ from stack locations in the set $L_s$ {(``Stack locations'' in Fig. \ref{fig:constructs-memorymodel})} to typed values, denoted as $v:\cd{t}$, where $v\in\cd{Val}$ is a value and $\cd{t}\in\mathcal{T}$ is its type.
Different stacks have disjoint location spaces in $L_s$ to ensure that the stacks
are local to their corresponding threads.
$L_s$ is partial order {under $\leq_s$} since only stack locations in {the} same thread are ordered {to denote
the creation order of local variables}.
Various strategies could be selected to implement $L_s$.
{For instance, a stack location could be a pair $(\mathit{tid},\cd{n})$, where $\mathit{tid}$ is a thread id and
$\cd{n}$ is a natural number to denote a local stack location of $\mathit{tid}$}.

The heap is defined as a set of blocks $H=\{B_1,\ldots,B_n\}$.
A block $B_i$ is 4-tuple $(\cd{b}_i,\cd{n}_i,m_i,\cd{t}_i)$, where $\cd{b}_i\in L_b$, $L_b$ is a set of block locations (``Block locations'' in Fig. \ref{fig:constructs-memorymodel}) and each block has a unique block location, $\cd{n}_i\in\mathbb{N}$ is the size of the block,
$m_i: [0, \cd{n}_i-1]\rightarrow \cd{Val}\times \mathcal{T}$ is a map from the offsets in the range $[0,\cd{n}_i-1]$ to the corresponding typed values {($[\cd{n}_1,\cd{n}_2]$ denotes the set of natural numbers $\cd{n}_1,\cd{n}_1+1,\ldots,\cd{n}_2$)}, and $\cd{t}_i\in \mathcal{T}$ is the block's type. A heap location is defined as a pair  $(\cd{b},\cd{m})$ (``Heap locations'' in Fig. \ref{fig:constructs-memorymodel}), where $\cd{b}$ is a block location and $\cd{m}$ is the offset within the block $\cd{b}$.

The definitions of  stacks and  heaps are capable of storing values of various types, such as primitive types {(integers, boolean values, among others)}, pointers, arrays, product types, and sum types. Heap blocks storing values of primitive types or pointers have {size 1}. {Blocks storing arrays or values of product types have size greater or equal than $1$, according to the number of elements the types are composed of, and two elements in the case of sum types: one for the value itself and another to indicate the constructor
that the sum type selects to construct the value.  Fig. \ref{fig-memorylayout} illustrates the stacks and heap created by the program on the left in the figure.}
Three variables $a,b,t$ created at lines 3, 4, 5 have the corresponding stack locations $s_0,s_1,s_2$, respectively.
Line 1 and 2 define a \cd{struct} type and an \cd{enum} type, respectively.
Line 3 creates a block to store an array of the type \cd{[i32:3]} (a 32 bits integer array of the size 3).
The location, size, and type of the block are \cd{b1}, \cd{3}, and \cd{[i32:3]} respectively.
The offsets 0, 1, 2 store the values 1, 2, 3 of type \cd{i32}, respectively.
Line 4 creates two  blocks \cd{b2} and \cd{b3}. The block \cd{b2} stores the value of the type \cd{L} defined by the constructor \cd{L} in the enumeration type \cd{B}.
{The symbol \cd{own(b1)} denotes a value storing the address of the block \cd{b1}.}
The block \cd{b3} stores the value of the type \cd{B}, which has two constructors: \cd{Empty} and \cd{L}.
{The value ``$1$'' at the offset $0$ in the block \cd{b3} indicates
	the constructor \cd{L} is selected, otherwise if it is $0$ then the constructor \cd{Empty} is selected.}
Line 5 creates the block \cd{b4} storing the value of type \cd{T}.

\begin{figure}
	\center
	\includegraphics[scale=0.48]{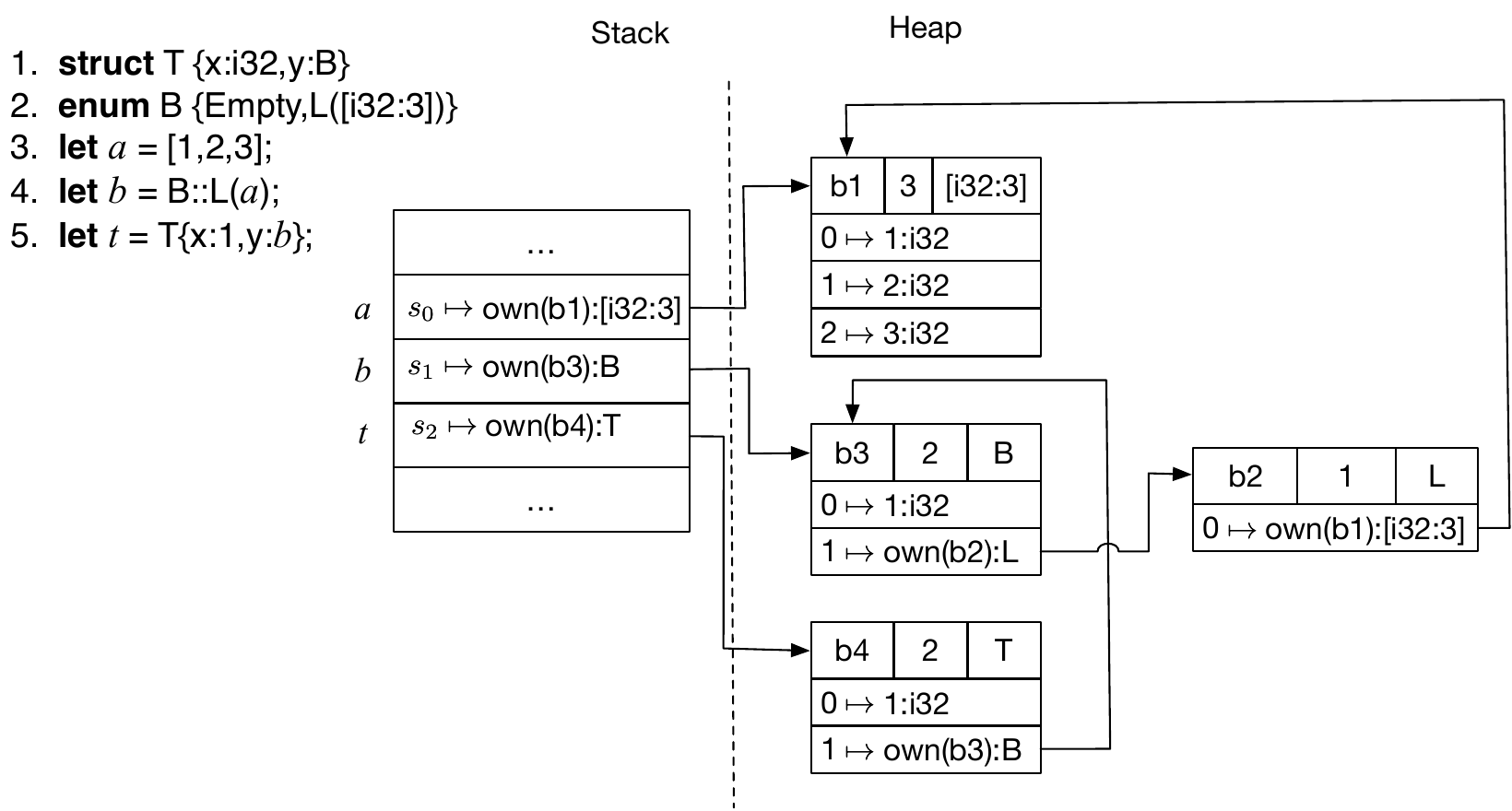}
	\caption{{An example of the stacks and heap}}
	\label{fig-memorylayout}
	\vspace{-0.5cm}
\end{figure}

In order to access memory locations, paths and aliases are introduced (Paths, Aliases in Fig. \ref{fig:constructs-memorymodel}).
A path could be a stack location, a {dereference}, or a field.
It can be used to access a memory block with an alias that refers to the block.
{
Consider the following code ``\cd{struct P\{x:i32,y:i32\} let v = P\{x:1,y:2\}; let z = \& v; }'',
if we want to access the field \cd{x} of \cd{v} with the alias \cd{z} then
 the path is $(* \cd{z}).\cd{x}$.
 Assume the stack location of \cd{b} is \cd{s} and the field \cd{x} corresponds
 to the offset $0$.
 The path in the memory model is $\cd{(* s)}.0$.
}
Each path $p$ has an alias, which is defined as:
{
	$
	\cd{alias}(p)\triangleq
	\begin{cases}
	p & p = \cd{s}\text{ or }p =\cd{s}.\cd{n}_1\ldots\cd{n}_m\\
	\cd{alias}(p') & {p = * p'\text{ or }p=(*p').\cd{n}}
	\end{cases}.
	$}
Roughly speaking, the alias of a path is the term innermost the path without dereferences.
An  alias of a block itself is also a memory location.
For instance, $s_1.1$ corresponds to the heap location {$(\cd{b3},1)$} in Fig. \ref{fig-memorylayout},
{which refers to the block \cd{b2}}.
{
  We define a predicate $\cd{des}(\cd{a}',\cd{a})$ for two aliases as $\cd{des}(\cd{a}',\cd{a})\triangleq\exists \cd{m}\geq 0.\cd{a}'=\cd{a}.\cd{n}_1.\ldots.\cd{n}_\cd{m}$. 
	It means that $\cd{a}'$ is a field or nested field of $\cd{a}$.
	For instance, in Fig. \ref{fig-memorylayout}, $\cd{s}_2.1$ is a field of $\cd{s}_2$,
	$\cd{s}_2.1.1$ if a field of $\cd{s}_2.1$ and the nested field of $\cd{s}_2$.
	Note that, $\cd{des}(\cd{a},\cd{a})=\emph{true}.$
	This is introduced for fine-grained borrowings.
	For a borrowing relation $\cd{a}\rightarrow_b\cd{a}'$, if $\cd{des}(\cd{a}'',\cd{a}')$ then $\cd{a}\rightarrow_b \cd{a}''$.
}

The third element in the configuration $mem$ is a  finite partial map $\mathcal{P}:\mathcal{A}\xrightharpoonup{\emph{fin}} \textit{Tim}\times \textit{Tim}$ from aliases to pairs of timestamps,
which stores the latest timestamps
at which an alias was used to {read and write, respectively.}
For instance, $\cd{s}\mapsto (\cd{ts}_1,\cd{ts}_2)\in \mathcal{P}$ means that
the latest reading and writing using $\cd{s}$ happened at $\cd{ts}_1$ and
$\cd{ts}_2$, respectively.  
This information is used
to check whether a reading or writing is disabled by other references.

Finally, $ms:L_m\xrightharpoonup{fin} \mathbb{N}\times\mathbb{N}$ is a finite partial  map from memory locations (``Memory locations'' in Fig. \ref{fig:constructs-memorymodel}) to pairs of natural numbers, used
to detect data races with respect to sequential consistency, where $l \mapsto(\cd{n}_1,\cd{n}_2)\in ms$ represents that
there are $\cd{n}_1$ and $\cd{n}_2$ reading and writing simultaneous accesses from/to the location $l$. The idea is inspired by Jung and others \cite{Ralf2018}.

{
At the end of this subsection, we introduce some notations for $mem=(S,H,\mathcal{P},ms)$.
For a partial map $M$, $M(k)$ denotes value of  the key $k$ in the map.
If $k$ has no value in $M$ then $M(k)$ is undefined.
The notation $M[k\leftarrow v]$ denotes a new map obtained from $M$
by replacing the value of $k$ with $v$, which is defined as:
$M[k\leftarrow v](k')\triangleq
	\begin{cases}
	v & k= k'\\
	M[v] & k\neq k'\\
	\end{cases}.$
Let $B=(\cd{b},\cd{n},m,\cd{t})$ be a block in $H$ and $l=(\cd{b},\cd{n}'),0\leq \cd{n}'< \cd{n}$.
$H(l)\triangleq m(\cd{n}')$ denotes the value stored in the heap location $l$. 
$H[l\leftarrow v]\triangleq (H\backslash \{B\})\cup\{(\cd{b},\cd{n},m[\cd{n}'\leftarrow v],\cd{t})\}$ 
denotes a new heap by replacing the value in $l$ with $v$.
Moreover, we use 
$mem(l)=v$ to denote $S(l)=v\lor H(l)=v$ and
$mem[l\leftarrow v]$ to denote a new memory configuration in which $S$ is replaced with $S[l\leftarrow v]$ or $H$ is replaced with $H[l\leftarrow v]$ when $l$ is a stack or a heap location, respectively.}

\subsection{Values of Memory Model}
\label{subsect-pointers}

We now introduce the kinds of values used in the memory.
Generally speaking, there are three kinds of values: primitive values, pointers, and $\bot$ denoting the uninitialization of a memory location.
Primitive values include integers, floating-pointer values, boolean values, characters, and strings, which 
are standard.
Here we focus on the elaboration of pointer values (``Pointer Values''  in Fig. \ref{fig:constructs-memorymodel}).
Pointer values consist of the following $3$ kinds:
\begin{itemize}
	\item {Own} pointers (\cd{own}(b)) to indicate the ownership of a block \cd{b}, an alias holds an \cd{own} pointer means it owns the block.
	\item Shared and mutable reference values.
	The shared (resp. mutable) reference  values have two forms $\cd{shr}(lt,p)$ and $\cd{shr}(p)$ (resp. $\cd{mut}(lt,p)$ and $\cd{mut}(p)$). The reason is that lifetimes are transparent for users using the model, whilst in {the} memory, a reference should be attached a lifetime for {OBS checking}.
	\item Unsafe pointers ($\cd{raw}(l)$), which are raw pointers to memory locations. Unsafe pointers belong to the unsafe features of Rust. 
\end{itemize}
Own pointers, shared and mutable reference values are safe pointers.
{Rust distinguishes safe and unsafe pointers to decide whether to carry out OBS invariant checking.}

{
The values $\cd{shr}(p)$ and $\cd{mut}(p)$ create references to $\cd{alias}(p)$.
For instance, we can write a borrowing like ``\cd{let x = \& (*v); }'' in Rust.
Assume the stack location of $\cd{v}$ is $\cd{s}$ then the reference value to be assigned to $\cd{x}$ is 
$\cd{shr}(* \cd{s})$. Here the path is $*\cd{s}$ and its alias is \cd{s}.
For two aliases $\cd{a}$ and ${\cd{a}'}$, if 
$\cd{a}$ holds a shared (resp. mutable) reference value to $\cd{a}'$, i.e., $\cd{shr}(lt,p)\land \cd{alias}(p)=\cd{a}'$ (resp. $\cd{mut}(lt,p)\land \cd{alias}(p)=\cd{a}'$) in $mem$, then we
also write $\cd{a}\rightarrow_s\cd{a}'\in mem$ (resp. $\cd{a}\rightarrow_m\cd{a}'\in mem$) and
the notation $\mathit{lifetime}(\cd{a},mem)$ denotes $lt$ in the values.
Moreover the notation $\cd{a}_1\rightarrow_b^* \cd{a}_\cd{n}\in mem$ denotes that there exists a sequence $\cd{a}_1, \cd{a}_2,\ldots,\cd{a}_{\cd{n}-1},\cd{a}_\cd{n}$ for $\cd{n}\geq 1$ such that $\cd{a}_i\rightarrow_b \cd{a}_{i+1}\in mem,~1\leq i <\cd{n}$,
where $\rightarrow_b$ denotes either $\rightarrow_m$ or $\rightarrow_s$.
Note that $\cd{a}\rightarrow_b^* \cd{a}\in mem$.
}

{
	The lifetime  $\cd{ts}_1\sim\cd{ts}_2$ in a reference value
     indicates that the reference is created at the timestamp $\cd{ts}_1$ and the last timestamp it is used to read or write is $\cd{ts}_2$. 
	Note that,
	In Fig. \ref{fig:constructs-memorymodel}, 
	the meta-variable $lt$ is called \emph{local lifetimes}.
	The reason is that
	for a reference $a$, 
	it is \emph{in use} if it is used to read, write, or 
	\emph{borrowed} by other references.
	The local lifetime is the span only from its creation to the timestamp that it is used to read or write.
	Therefore the exact lifetime of a reference in $mem$, denoted as $\mathcal{L}(\cd{a},mem)$,
	is defined as $\mathcal{L}(\cd{a},mem)=\bigcup_{\cd{a}'\in\{\cd{a}''\mid \cd{a}''\rightarrow_b^* \cd{a}\in mem\}}{\mathit{lifetime}(\cd{a}',mem)}$.
	It is the union of all local lifetimes of $a'$ such that $a'\rightarrow^*_b a\in mem$.
	%
	%
}

\subsection{Lifetime-Free Memory Operation Interfaces}
\label{subsection-Operationinterfaces}

The memory operation interfaces (``Memory Operations'' in Fig. \ref{fig:constructs-memorymodel}) are lifetime-free, which means that the parameters of memory operations have no lifetimes.
This design aims at abstracting lifetime information from the memory interfaces to enable reusability, since
other languages may not have the notion of lifetimes. Memory operations include:
\begin{itemize}
	\item Allocation ($\cd{alloc}(\cd{n},\cd{t})$), allocates a new memory block of size \cd{n} for storing the value of the type \cd{t}. Free (\cd{free}(\cd{b})), deallocates the memory block \cd{b}.
	\item Raw reading and writing ($\cd{rawRead}(l)$, $\cd{rawWrite}(l,v)$) {provide} non-atomic reading and writing without integrating OBS invariant checking. 
	\item Atomic reading and writing ($\cd{aRead}(l)$, $\cd{aWrite}(l,v)$) {provide} atomic reading and writing without integrating OBS invariant checking.
	\item Reading and Writing ($\cd{read}(p)$, $\cd{write}(p,v)$) {provide} reading and writing with the integration of OBS invariant checking.
	\item {Lvalue ($\cd{lv}(p)$), computes the  Lvalue of the path $p$. Lvalue of a path is the memory location
	 identified by the path and Rvalue is the value stored in the Lvalue of the path.}
\end{itemize}

Here we need to restrict that  if the parameter \cd{v} in either \cd{rawWrite}, \cd{aWrite}, or \cd{write} is a reference value then
it can only be $\cd{shr}(l)$ or $\cd{mut}(l)$, i.e., no lifetimes. 
Actually, lifetime computation is hidden in the implementation of memory operations, which will be introduced in the following. 

{The semantics is defined by two kinds of transition relations. The first one is $\cfg{mem,tm}^{\cd{ts}}\rmem\cfg{mem',tm'}^{\cd{ts}'}$. $tm$ and $tm'$ are terms that can be a value, a memory operation, or ``\cd{.}'' indicating that the operation is consumed. 
The symbols $\cd{ts}$ and $\cd{ts}'$ are the timestamps of the pairs and
satisfying $\cd{ts}\leq\cd{ts}'$.
The second one is $\cfg{mem,tm}^{\cd{ts}}\rmem\cd{stuck}$ indicating the semantics gets stuck.}
Some of $S,H,\mathcal{P},ms$ can be {omitted} in rules if they are not used  to make 
rules more concise.

\subsubsection{Operational Semantics for Allocation and Free}


%
Rule \refrule{Allocation} defines the semantics for $\cd{alloc}(\cd{n},\cd{t})$, where
$\cd{n}$ and $\cd{t}$ are the size and type of the new block, respectively. 
It creates a new block with a fresh block location ($\emph{fresh}(H)$) and adds the new block to the heap $H$. 
Moreover, it also initializes $ms$ for the new block.
$\cd{initBlk}(\cd{n})$ denotes the map $\{0\mapsto\bot,\ldots ,\cd{n}-1\mapsto \bot\}$.
$\cd{initMS(b,n)}=\{\cd{(b,0)}\mapsto (0,0),\ldots,\cd{(b,n-1)}\mapsto (0,0)\}$.
\refrule{Free} removes a block from $H$.
The set $bLoc(H)$ denotes all the block locations used in $H$, i.e., for any {$\cd{b}\in bLoc(H)$}, there is a block $(\cd{b}, \cd{n},m,\cd{t})$ in $H$. ``$\_$'' matches anything. {The timestamps are increased
by $\mathit{Su}(\cd{ts})$.}
{
\TwoColumn{
	\ruleEnv{Allocation\label{Allocation}}{
		$\cd{b}=\emph{fresh}(H)$\quad$H'=H\cup(\cd{b},\cd{n},\cd{initBlk}(\cd{n}),\cd{t})$\\
		$ms'=ms\cup
		\cd{initMS(\cd{b},\cd{n})}$\\
		\hline
		$\cfg{(H,ms),\codec{alloc}(\cd{n},\cd{t})}^\cd{ts}\rmem\cfg{(H',ms'),\cd{own(b)}}^{\mathit{Su}(\cd{ts})}$
	}}{
	\ruleEnv{Free\label{Free}}{
		$\cd{b}\in bLoc(H)\quad H'=H\backslash \{(\cd{b},\cd{n},\_,\_)\}\quad\cd{n}\geq 0$\\
		$ms'=ms\backslash \{\cd{(b,0)}\mapsto\_,\ldots,\cd{(b,n-1)}\mapsto\_\}$\\
		\hline
		$\cfg{(H,ms),\cd{free}(\cd{b})}^\cd{ts}\rmem{}\cfg{(H',ms'),.}^{\mathit{Su}(\cd{ts})}$
}}
}

\subsubsection{Operational Semantics for Non-Atomic Raw Reading and Writing}
%

The rules for \cd{rawWrite} and \cd{rawRead}
are non-atomic {writing and reading} that are directly applied to raw pointers. The execution of a raw operation will get stuck
under data races.
For a location $l$ and $l\mapsto(\cd{n}_1,\cd{n}_2)\in ms$, data races are defined as (1) $(\cd{n}_1+\cd{n}_2\geq 2)\land (\cd{n}_2\geq 1)$, i.e., there are at least two threads accessing the location and at least one of them writes the location.
In order to simulate non-atomic operations, both raw reading and writing are decomposed into
two steps that can be interleaved.

The raw writing is defined by Rule \refrule{RawWrite} and \refrule{RawWrite$'$}. It is a two step operation where the first step modifies $ms$ and translates
$\cd{rawWrite}(l,v)$ to $\cd{rawWrite}'(l,v)$ and $ms(l)=(0,0)$ ensures that the writing will not cause a data race. 
The second step writes $v$ in the heap location $l$, and resets $ms$. The two steps can be interrupted by other threads. 
The semantics of raw reading is similar to the raw writing, which is defined by Rule \refrule{RawRead} and \refrule{RawRead$'$}.
{Timestamps are not increased, since they are not for safe pointers.}
We still need the semantics for detecting race conditions.
Rule \refrule{Race-RawRead} defines the semantics for the race conditions of the reading. It evolves
into the stuck state.
The race semantics for other operations are similar.

\TwoColumn{
	\ruleEnv{RawWrite\label{RawWrite}}{
		$ms(l)=(0,0)$\quad
		$ms'=ms[l\leftarrow(0,1)]$ \\
		\hline
		$\cfg{{(H,ms)},\cd{rawWrite}(l,v)}^\cd{ts}$\quad\quad\quad\quad\quad\\
		\quad\quad\quad\quad$\rmem{}\cfg{(H,ms'),\cd{rawWrite}'(l,v)}^\cd{ts}$
}\quad\quad}
{
	\ruleEnv{RawWrite$'$\label{RawWrite$'$}}{
		$ms(l)=(0,1)$\quad
		$H'=H[l\leftarrow v]$\\
		$ms'=ms[l\leftarrow(0,0)]$ \\
		\hline
		$\cfg{(H,ms),\cd{rawWrite}'(l,v)}^\cd{ts}\rmem{}\cfg{(H',ms'),.}^\cd{ts}$
}}

\TwoColumn{
	\ruleEnv{RawRead\label{RawRead}}{
		$ms(l)=(\cd{n},0)$\quad$\cd{n}\geq0$\quad
		$ms'=ms[l\leftarrow(\cd{n}+1,0)]$ \\
		\hline
		$\cfg{(H,ms),\cd{rawRead}(l)}^\cd{ts}$\quad\quad\quad\quad\quad\\
		\quad\quad\quad\quad\quad$\rmem{}\cfg{(H,ms'),\cd{rawRead}'(l,v)}^\cd{ts}$
}\quad\quad}
{
	\ruleEnv{RawRead$'$\label{RawRead$'$}}{
		$ms(l)=(\cd{n},0)$\quad
		$H(l)=v$\quad$v\neq\bot$\\
		$\cd{n}\geq 1$\quad
		$ms'=ms[l\leftarrow(\cd{n}-1,0)]$ \\
		\hline
		$\cfg{(H,ms),\cd{rawRead}'(l)}^\cd{ts}\rmem{}\cfg{(H,ms'),v}^\cd{ts}$
}}

\TwoColumn{
\ruleEnv{Race-RawRead\label{Race-RawRead}}{
	$mem=(S,H,\mathcal{P},ms)$\quad$ms(l)=(\cd{n}_1,\cd{n}_2)\land \cd{n}_2>0$\\
	\hline
	$\cfg{mem,\cd{rawRead}(l)}^\cd{ts}\rmem \cd{stuck}$
}}{
\ruleEnv{Concur-Stuck\label{Concur-Stuck}}{
	$\cfg{mem,t_1}^\cd{ts}\rmem \cd{stuck}$ or $\cfg{mem,t_2}^{\cd{ts}}\rmem \cd{stuck}$\\
	\hline
	$\cfg{mem,t_1\mid\mid t_2}\leadsto_c \cd{stuck}$
}
}

\TwoColumn{
\ruleEnv{Concur-1\label{Concur-1}}{
	$\cfg{mem,t_1}^\cd{ts}\rmem\cfg{mem',t}^\cd{ts}$\\
\hline
$\cfg{mem, t_1\mid\mid t_2}\leadsto_c\cfg{mem',t\mid\mid t_2}$
}}{
\ruleEnv{Concur-2\label{Concur-2}}{
	$\cfg{mem,t_2}^\cd{ts}\rmem\cfg{mem',t}^\cd{ts}$\\
	\hline
	$\cfg{mem, t_1\mid\mid t_2}\leadsto_c\cfg{mem',t_1\mid\mid t}$
}}

{
For instance, 
	we consider the concurrent execution of two operations $\cd{rawRead}(l)$ and $ \cd{rawWrite}(l,v)$ under a concurrent semantics based on interleaving, whose grammar is
	$con::=\mathit{tm}\mid\mid \mathit{tm}$, where $\mathit{tm}$ can be 
	a memory operation, a value, or ``.'' The semantics  is defined by the Rule \refrule{Concur-1}, \refrule{Concur-2}, and \refrule{Concur-Stuck}, with the relation $\leadsto_c$ (The concurrent semantics is only for illustration here. RustSEM concurrent semantics
	is implemented in CL level.}

{
	Assume the initial memory configuration satisfies  $ms(l)=(0,0)$ and $H(l)\neq\bot$,
there are $4$ possible execution sequences as follows:
}

{
(Sequence 1) $\cd{rawRead}(l);~~\cd{rawRead}'(l);~~$
$\cd{rawWrite}(l,v);~~\cd{rawWrite}'(l,v)$;}

{
(Sequence 2) $\cd{rawWrite}(l,v);$ $\cd{rawWrite}'(l,v);~~$
		$\cd{rawRead}(l);~~\cd{rawRead}'(l)$;}

{
(Sequence 3) $\cd{rawRead}(l);~~\cd{rawWrite}(l,v);~~\cd{stuck}$;}

{
(Sequence 4) $\cd{rawWrite}(l,v);~~\cd{rawRead}(l);~~\cd{stuck}$.}

{
Sequences (1) and (2) are safe, but Sequence (3) and (4) are unsafe.
For instance, in Sequence (4), 
after executing $\cd{rawWrite}(l,v)$, we have that $ms(l)=(0,1)$, which makes $\cd{rawRead}(l)$ get stuck (Rule \refrule{Race-RawRead}).
We will not present atomic reading and writing without OBS invariant checking as their semantics rules are trivial.
}

\subsubsection{Operational Semantics for Reading and Writing with OBS Checking}

The reading and writing operations for the safe pointers need to maintain the invariants of OBS.
The invariants correspond to the well-formed memory configurations, which are defined as follows. 

\begin{definition}[Well-formed memory configurations]\label{def-wellformlayouts}
	Let $mem=(S,H,\mathcal{P},ms)$ be a memory configuration.
	It is well-formed, denoted as $\cd{wellform}(mem)$, iff it satisfies the following invariants.
	
	\begin{enumerate}
		\item $\forall \cd{b},l,l'.(mem(l)=mem(l')=\cd{own}(\cd{b}))\Longrightarrow l= l'$.
		\item $\forall \cd{a},\cd{a}_1,\cd{a}_2. \cd{a}_1\rightarrow_m \cd{a}\in mem\land
		\cd{a}_2\rightarrow_b \cd{a}'\in mem\land\cd{des}(\cd{a},\cd{a}')\Longrightarrow {\mathcal{L}(\cd{a}_1,mem)\cap \mathcal{L}(\cd{a}_2,mem)=\emptyset}$. 
		\item $\forall \cd{a},\cd{a}'. \cd{a}\rightarrow_s\cd{a}'\in mem\Longrightarrow (\forall \cd{a}_1.\cd{des}(\cd{a}_1,\cd{a}')\land \mathcal{P}(\cd{a}_1)=(\cd{ts},\cd{ts}')\Longrightarrow  \cd{ts}'\notin \mathcal{L}(\cd{a},mem))$.
		\item $\forall \cd{a},\cd{a}'. \cd{a}\rightarrow_m\cd{a}'\in mem\Longrightarrow (\forall \cd{a}_1.\cd{des}(\cd{a}_1,\cd{a}')\land \mathcal{P}(\cd{a}_1)=(\cd{ts},\cd{ts}')\Longrightarrow  \cd{ts}\notin \mathcal{L}(\cd{a},mem)\land \cd{ts}'\notin \mathcal{L}(\cd{a},mem)$.
		\item $\forall l.ms(l)=(n_1,n_2)\Longrightarrow \neg ((n_1 + n_2 \geq 2)\land (n_2>1))$.
	\end{enumerate}
\end{definition}

Invariant $(1)$ ensures that no block is owned by more than one locations.
Invariant $(2)$ {ensures} that an alias cannot be borrowed by two references simultaneously with one of them being mutable.
Invariant $(3)$ ensures if an alias is borrowed by a shared reference then
its latest writing {timestamp} cannot be in the lifetime of the reference. 
Invariant $(4)$ ensures if an alias is borrowed by a mutable reference then both its latest reading and writing timestamps cannot be in the lifetime of the reference.
{The last invariant ensures no data race.}

Compared with well-formed OBS graphs, 
well-formed memory configurations also
contain \emph{unique owner} 
and \emph{lifetime disjoint} invariants,
but no \emph{lifetime inclusion} invariant.
Lifetime inclusion invariant will be  maintained by the semantics rules directly.
Invariants (4) and (5) of well-formed memory configurations correspond to 
{the} permission functions, but not exactly since it only disable the latest reading and writing.
We will prove that it enforces memory operations to follow the permission functions in Subsection \ref{subsection-refinement}.
Data race is not considered in {the} well-formed OBS graphs, since OBS graphs only consider about safe pointers

\paragraph{The semantics of Lvalue}

The lvalue of a path is the location identified by the path that is to be written or read.
The lifetimes of references accessed during
the computation of \cd{lv} 
{
should be extended since the references are used.
The result of $\cd{lv}(p)$ is pair $(l,wp)$ where $l$ is the Lvalue
and $wp$ is a boolean value indicates whether $p$ is permitted to write.
A path is permitted to write iff the borrowing links ``$\cd{alias}(p)\rightarrow_m \cd{a}\ldots\cd{a}'\rightarrow_m\cd{x}$''} from the alias $\cd{alias}(p)$ to the owner $\cd{x}$
are all mutable borrowings.

The first rule for Lvalue is \refrule{Lv-Deref}, which computes the Lvalue of $*p$, where $p$ is a path.
We use $\cd{ref}(p')$ (resp. $\cd{ref}(lt,p')$) to denote a reference, which can be either a shared reference $\cd{shr}(p')$ (resp. $\cd{shr}(lt,p')$) or a mutable reference $\cd{mut}(p')$ (resp. $\cd{mut}(lt,p')$).
\begin{itemize}
	\item Premise (1) computes the {Lvalue} of $p$, which is $l$. {Lvalue semantics does not increase
	the timestamp as it is one of the sub-steps for reading and writing.}
	\item {Premise (2) checks whether the location is being written by a non-atomic operation.
	$\llbracket mem_1\rrbracket.ms$ denotes the element $ms$ in the memory configuration $mem_1$. 
	It requires that the number of writings is $0$ and the number of readings is greater than or equal to $0$.
	Therefore it could be used to check the data races that a non-atomic writing is carrying out but interleaved
	by safe operations.}
%
	\item {Premise (3) requires that the location ${l}$ must be a shared or mutable reference, since we can only
	dereference a reference. It is a shorthand for $\exists p', lt. mem_1(l)=\cd{ref}(lt,p')$.}
	\item {As the reference $l$ is used, its lifetime is extended to the  timestamp \cd{ts}  (Premise (4)) by the function \cd{extLT}, which is defined as follows.}
	\[\small
	\cd{extLT}(mem,l,\cd{ts}){\triangleq}\begin{cases}
	mem[l\leftarrow \cd{shr}(\cd{ts}_1\sim\cd{ts}, p')] & mem(l)=\cd{shr}(\cd{ts}_1\sim\cd{ts}_2, p')\\
	mem[l\leftarrow \cd{mut}(\cd{ts}_1\sim\cd{ts}, p')] & mem(l)=\cd{mut}(\cd{ts}_1\sim\cd{ts}_2, p')
	\end{cases}
	\]
	\item Since the referent $p'$ is a path, its {Lvalue} needs to be further computed (Premise (5)).
	\item Premise (6) ensures the memory configuration is still well-formed after the computation.
	\item Premise (7) checks whether $l$ is a mutable reference (The notation $e?v_1,v_2$ denotes that if $e$ is true then $v_1$ otherwise $v_2$). 
\end{itemize}

The resulting permission is decided by $wp\land wp_1\land wp_2$, i.e., whether all references
accessed by the semantics rule are mutable references.
{
Rule \refrule{Lv-Location} and \refrule{Lv-Field} compute the Lvalues of a location (which is itself) and  a field, respectively.}
{
\RuleEnv{Lv-Deref\label{Lv-Deref}}{
	(1) $\cfg{mem,\cd{lv}(p)}^\cd{ts}\rmem\cfg{mem_1,(l,wp_1)}^{\cd{ts}}$\quad
	(2) $\llbracket mem_1 \rrbracket. ms(l)=(\cd{n},0)\land \cd{n}\geq 0$\\
	(3) $mem_1(l)=\cd{ref}(lt,p')$\quad
	(4) $mem_2=\cd{extLT}(mem_1,l,\cd{ts})$\quad
	(5) $\cfg{mem_2,\cd{lv}(p')}^{\cd{ts}}\rmem\cfg{mem_3,(l',wp_2)}^{\cd{ts}}$\\
	(6) $\cd{wellform}(mem_3)$\quad
	{
	(7) $wp=(mem_1(l)=\cd{mut}(lt,p')?\emph{true}:\emph{false})$}\\
	\hline
	$\cfg{mem,\cd{lv}(* p)}^{\cd{ts}}\rmem\cfg{mem_3,(l',wp\land wp_1\land wp_2 )}^{\cd{ts}}$}}
\noindent
{
\TwoColumn{
	\ruleEnv{Lv-Location\label{Lv-Location}}{
		$\cfg{mem,\cd{lv}(l)}^\cd{ts}\rmem{} \cfg{mem',(l,\emph{true})}^{\cd{ts}}$
	}
}{
	\ruleEnv{Lv-Field\label{Lv-Field}}{
		(1) $\cfg{mem,\cd{lv}(p)}^\cd{ts}\rmem\cfg{mem_1,(l,wp)}^{\cd{ts}}$\\
		(2) $mem_1(l)=\cd{own}(\cd{b})$\quad (3) $\cd{b}\in bLoc(mem_1)$\\
		\hline
		$\cfg{mem_,\cd{lv}(p.\cd{n})}^\cd{ts}\rmem{}\cfg{mem_1,((\cd{b},\cd{n},),wp)}^{\cd{ts}}$}
}}

For instance, assume in the memory, we have $\cd{s}_1\mapsto \cd{mut}(1\sim 2, \cd{s}_2),\cd{s}_2\mapsto \cd{own}(b)$ { and $ms(\cd{s}_1)=(0,0), ms(\cd{s}_2)=(0,0)$.}
If we try to compute $\cd{lv}(*\cd{s}_1)$, 
Rule \refrule{Lv-Deref} works as follows.
\begin{itemize}
	\item Premise (1) computes $\cfg{mem,\cd{lv}(\cd{s}_1)}$
	and {gets} $(\cd{s}_1,\emph{true})$ by Rule \refrule{Lv-Location}.
	\item {Premise (2) and (3)  are true as $ms(\cd{s}_1)=(0,0)$ and $\cd{s}_1$ holds $\cd{mut}(1\sim 2,\cd{s}_2)$.}
	\item Premise (4) extends its lifetime to the current timestamp (assume it is $4$). Thus value of $\cd{s}_1$
	is updated as $\cd{mut}(1\sim 4,\cd{s}_2)$ now.
	\item Premise (5) further computes the lvalue of $\cd{lv}(\cd{s}_2)$, which is $(\cd{s}_2,true)$. Therefore $\cd{lv}(*\cd{s}_1)$ is $\cd{s}_2$.
	\item The writing permission is true as $\cd{s}_1$ a mutable reference and the writing permission of both $\cd{lv}(\cd{s}_1)$ and $\cd{lv}(\cd{s}_2)$ are true.
\end{itemize}

\paragraph{The semantics of writing}
The semantics of $\cd{write}(p,v)$, writes the value $v$ to the Lvalue of $p$.
Rule \refrule{Write-Ref} defines the semantics for writing a reference value $\cd{ref}(p')$
to the {Lvalue} of the path $p$.
The semantics is elaborated as follows.
{
\begin{itemize}
	\item Premise (1) computes the Lvalue of the path $p$ by $\cd{lv}(p)$ and requires the permission to write.
	\item Premise (2) ensures no reading or writing to the location $l$,
	\item Premise (3) writes the value $v$ to the location $l$.
	The reference value should be attached with the local lifetime $\cd{ts}\sim\cd{ts}$, i.e., the reference starts to have a lifetime.
	\item Premise (4) updates the latest writing timestamp of  $p$'s alias by the function
	\cd{addWrite}.
	Assume $mem=(S,H,\mathcal{P},ms)$, 
	$\cd{addWrite}$ is defined as:
	$
	\cd{addWrite}(mem,\cd{a},\cd{ts})\triangleq (S,H,\mathcal{P}[\cd{a}\leftarrow (\cd{ts}_1,\cd{ts})],ms)~\cd{if }\mathcal{P}(\cd{a})=(\cd{ts}_1,\cd{ts}_2).
	$
	\item Premise (5) ensures that the memory configuration is well-formed after the writing.
\end{itemize}}
%

%

{
\RuleEnv{Write-Ref\label{Write-Ref}}{
	(1) $\cfg{mem,\cd{lv}(p)}^{\cd{ts}}\rmem\cfg{mem_1,(l,\emph{true})}^{\cd{ts}}$\quad
	%
	(2) $\llbracket mem_1 \rrbracket. ms(l)=({0},0)$\\
    (3) $mem_2=mem_1[l\leftarrow \cd{ref}(\cd{ts}\sim\cd{ts},p')]$\quad
	(4) $mem_3=\cd{addWrite}(mem_2,\cd{alias}(p),\cd{ts})$
\quad (5) $\cd{wellform}(mem_3)$\\
	\hline
	$\cfg{mem,\cd{write}(p,\cd{ref}(p'))}^{\cd{ts}}\rmem\cfg{mem_3, .}^{\mathit{Su}(\cd{ts})} $
}}

{
\TwoColumn{
\ruleEnv{Write-Own\label{Write-Own}}{
	(1) $\cfg{mem,\cd{lv}(p)}^{\cd{ts}}\rmem\cfg{mem_1,(l,\emph{true})}^{\cd{ts}}$\\
	%
	(2) $\llbracket mem_1 \rrbracket. ms(l)=({0},0)$\\
	(3) $mem_2=mem_1[l\leftarrow \cd{own}(b)]$\\
	(4) $mem_3=\cd{addWrite}(mem_2,\cd{alias}(p),\cd{ts})$\\
	(5) $\cd{wellform}(mem_3)$\quad
	(6) $\cd{b}\in bLoc(mem)$\\
	\hline
	$\cfg{mem,\cd{write}(p,\cd{own}(\cd{b}))}^{\cd{ts}}\rmem\cfg{mem_3, .}^{\mathit{Su}(\cd{ts})} $
}\quad
}{
\ruleEnv{Write-Primitive\label{Write-Primitive}}{
	(1) $\cfg{mem,\cd{lv}(p)}^{\cd{ts}}\rmem\cfg{mem_1,(l,\emph{true})}^{\cd{ts}}$\\
	(2) $v\in SV$\quad
	%
	(2) $\llbracket mem_1 \rrbracket. ms(l)=({0},0)$\\
	(3) $mem_2=mem_1[l\leftarrow v]$\\
	(4) $mem_3=\cd{addWrite}(mem_2,\cd{alias}(p),\cd{ts})$\\
	\hline
	$\cfg{mem,\cd{write}(p,v)}^{\cd{ts}}\rmem\cfg{mem_3, .}^{\mathit{Su}(\cd{ts})} $
}}}

{
Rule \refrule{Write-Own} is the semantics for writing an owner to a memory location.
Rule \refrule{Write-Primitive} is the semantics for writing primitive values, such as integers and boolean values.}

.

\paragraph{The semantics of reading} The operation $\cd{read}(p)$ reads a value from the Lvalue of a path $p$. \refrule{Read-Ref} defines the semantics for reading a reference.
{
\begin{itemize}
	\item Premise (1) computes the Lvalue of $p$, which is $l$ and Premise (3) reads the value from $l$,
	\item Premise (2) ensures no writing to the location. Premise (4) requires the value stored in $l$ should be a shared or mutable reference.
	\item Premise (5) updates the latest timestamp, at which the alias of $p$ is read, by the function \cd{addRead}, which is similar to \cd{addWrite}.
	\item Premise (6) extends the lifetime of $l$ as it is used and Premise (7) checks well-formedness..
\end{itemize}}
{
\RuleEnv{Read-Ref\label{Read-Ref}}{
	(1) $\cfg{mem,\cd{lv}(p)}^{\cd{ts}}\rmem\cfg{mem_1,(l,\_)}^{\cd{ts}}$\quad
	(2) $\llbracket mem_1\rrbracket.ms(l)=(\cd{n},0)\land \cd{n}\geq 0$\quad
	(3) $v=mem_1(l)$\\
	(4) $v=\cd{ref}(lt,p')$\quad
	(5) $mem_2=\cd{addRead}(mem_1,\cd{alias}(p),\cd{ts})$\\
	(6) $mem_3=\cd{extLT}(mem_1,l,\cd{ts})$\quad
	(7) $\cd{wellform}(mem_3)$\\
	\hline
	$\langle mem,\cd{read}(p)\rangle^{\cd{ts}}\rmem{}\langle mem_3, v \rangle^{\mathit{Su}(\cd{ts})}$ }
\RuleEnv{Read-Non-Reference\label{Read-Non-Reference}}{
	(1) $\cfg{mem,\cd{lv}(p)}^{\cd{ts}}\rmem\cfg{mem_1,(l,\_)}^{\cd{ts}}$\quad
(2) $\llbracket mem_1\rrbracket.ms(l)=(\cd{n},0)\land \cd{n}\geq 0$\quad
(3) $v=mem_1(l)$\\
	(4) $v$ is not a reference\quad
	(5) $v\neq\bot$\quad
	(7) if $v=\cd{own}(b)$ then $\cd{b}\in bLoc(mem)$\\ (6) $mem_2=\cd{addRead}(mem_1,\cd{alias}(p),\cd{ts})$\\
	\hline
	$\langle mem,\cd{read}(p)\rangle^{\cd{ts}}\rmem{}\langle mem_2,v \rangle^{\mathit{Su}(\cd{ts})}$  }}
Rule \refrule{Read-Non-Reference} defines the semantics of reading non-reference values.

{
\emph{Compared with weak memory models} \cite{DBLP:journals/jfp/JungKJBBD18, DBLP:journals/pacmpl/PodkopaevLV19},  RustSEM only concerns about sequential consistency and lies
in higher abstraction level without bridging the gap between OBS and specific hardware architectures. 
We could further refine our memory model by incorporating weak memory consistency in the future to strengthen its usability.}

\subsection{Refinement Relation between High-Level OBS and Memory Model}
\label{subsection-refinement}

{
	In this subsection, we prove the
 refinement relation between high-level abstraction of OBS and the memory model.
	We begin with the definition of safe sequences.}

{
	\begin{definition}
		Let $\pi= (mem_0,\cd{ts}_0),op_1,(mem_1,\cd{ts}_1),\ldots, (mem_{n-1}, \cd{ts}_{n-1}),op_n,$ $(mem_n,\cd{ts}_n)$ be an alternating sequence of memory configurations with timestamps and operations. For each pair $(mem,\cd{ts})$, $mem$ is a memory configuration and $\cd{ts}$ is the timestamp reaching $mem$.
		$\pi$ is a safe sequence iff 
		\begin{enumerate}
			\item $mem_0$ is an empty configuration, i.e., $S$, $\mathcal{P}, ms$ are empty maps and $H$ is an empty set, 
			\item $op_i,1\leq i\leq n$,  is one of the memory operations: 
			$\cd{alloc}(\cd{n}$, $\cd{t})$, $\cd{read}(p)$, $\cd{write}(p,v)$, $\cd{free}(\cd{b})$,
			\item $\cfg{mem_i,op_{i+1}}^{\cd{ts}_i}\rmem \cfg{mem_{i+1},re}^{\cd{ts}_{i+1}}$, for all $0\leq i < n$, where $re$ can only be a value if $\cd{op}_{i+1}$ is \cd{read} or \cd{alloc}, otherwise consumed ``.''.
		\end{enumerate}
\end{definition}}

{The memory configurations are the \emph{data refinement} of the high-level OBS graphs.
	For each memory configuration, it has an underlying OBS graph.}
{
	\begin{definition}
		\label{def-data-refinement}
		Let $\pi= (mem_0,\cd{ts}_0),op_1,(mem_1,\cd{ts}_1),\ldots, (mem_{n-1}, \cd{ts}_{n-1}),op_n,(mem_{n}, \cd{ts}_{n})$ be a safe sequence and $B$ be a block in $mem_i,(0\leq i \leq n)$, whose location is $\cd{b}$.
		The OBS graph $(V,B,E,\mathcal{F})$ of $mem_i$ is defined as:
		\begin{itemize}
			\item $V$ is a set of aliases in $mem_i$ that can access $B$,
			\item for any node $a$ in $V$, $
			E(a)=\begin{cases}
			(o,B) & \emph{if}~~mem(a)=\cd{own}(\cd{b})\\
			(s,a') & \emph{if}~~a\rightarrow_s a''\in mem_i\land\cd{des}(a',a'')\\
			(m,a') & \emph{if}~~a\rightarrow_m a''\in mem_i\land\cd{des}(a',a'')\\
			\end{cases}$
			\item $\mathcal{F}(a\rightarrow_b a')=\mathcal{L}(a,mem_i)$.
			$\mathcal{F}(a\rightarrow_o B)=\cd{ts}_j\sim\cd{ts}_i$ where for all $j\leq k\leq i$, $mem_k(a)=\cd{own}(\cd{b})$ and $mem_{j-1}(a)\neq\cd{own}(b)$.
		\end{itemize}
	\end{definition}
	It is now sufficient to present Theorem \ref{theorem-refinement}, which specifies the refinement relation between the high-level abstraction and the memory model.}

{
	\begin{theorem}
		\label{theorem-refinement}
		Let $\pi= (mem_0,\cd{ts}_0),op_1,(mem_1,\cd{ts}_1),\ldots, (mem_{n-1}, \cd{ts}_{n-1}),op_n,(mem_{n}, \cd{ts}_{n})$ be a safe sequence and $B$ be a block.
		Assume the OBS graph of $B$ in $mem_i~(0\leq i\leq n)$ is $G=(V,B,E,\mathcal{F})$ then we have
		\begin{enumerate}
			\item (Data refinement) $G$ is well-formed.
			\item (Operation refinement) For any two nodes $a,a'\in V\cup\{B\}$ such that $\mathcal{F}(a\rightarrow_* a')=lt$,
			for any operation $op_k~(0< k\leq n)$, which is executed at the timestamp $\cd{ts}\in lt$, 
			\begin{itemize}
				\item   if $op_k$ is an operation that reading $B$ by the alias $a$ then $R_{G}(a,\cd{ts})=\text{true}$,
				\item  if $op_k$ is an operation that writing $B$ by the alias $a$ then $W_{G}(a,\cd{ts})=\text{true}$.
			\end{itemize}
		\end{enumerate}
\end{theorem}}
{
	The proof is in Appendix \ref{append-thm-refinement}.
	Theorem \ref{theorem-refinement} shows that all operations by an alias $a$ follow the permission functions of $mem_i$'s OBS graph for $B$.}

\revision{
}

\section{CL Language and Translation Semantics}
\label{sec:rustsemacs}

In this section, we present the basic idea of CL language and the translation semantics, the supported features of Rust, and the efforts of developing RustSEM.
The semantics rules are not presented due to page limitation.
Actually, the core of RustSEM, OBS, is already formalized in the memory model.

The design of CL is to reduce Rust grammar to a simple core in order to avoid {redundant formalization}.
%
Table \ref{tab:correspondence-rust-cl} illustrates the Rust features supported by RustSEM (Column ``Rust features'')
and the corresponding grammars in CL (Column ``CL grammar'') for implementing the Rust features on the same row.
\begin{table}[t]
	\begin{center}
		\caption{Correspondence between Rust and CL features}
		\label{tab:correspondence-rust-cl}
		{\small
			\begin{tabular}[t]{@{}p{6.5cm}|p{6.5cm}}
				\toprule
				\centering \textbf{Rust features} &\quad\quad\quad\quad\quad\quad\quad\quad\textbf{CL grammar} \\ 
				\hline
				\midrule
				$\bullet$ primitive values
				$\bullet$ shared and mutable references \newline
				$\bullet$ casting pointers
				$\bullet$ arithmetic expressions\newline
				$\bullet$ print $\bullet$ assertion $\bullet$ move semantics
				$\bullet$ dereference\newline
				$\bullet$ field access $\bullet$ array access
				& 
				$e::=v \mid x\mid * e \mid e . e'\mid\cd{allocInit}(\cd{n},\vec{e} ,\cd{t})$
				
				\quad\quad$\mid e_1 + e_2\mid e_1 < e_2\mid e_1 \cd{\&\&}e_2\mid\ldots$
				
				\quad\quad$\mid \cd{move}(e)\mid\cd{\&}e\mid\cd{\& mut }e $
				
				\quad\quad$\mid  \cd{print}(e)\mid \cd{clskip}\mid \cd{assert}(e)$\\
				\hline
				$\bullet$ binding
				$\bullet$ execution block
				& 
				$binding::=\cd{let}~x=e~\cd{in}~\{e'\}$
				\\
				\hline
				$\bullet$ assignment &
				$assign::=e\cd{:=}e'$\\
				\hline
				$\bullet$ function ~$\bullet$ impl block
				$\bullet$ trait definition $\bullet$ trait impl
				
				$\bullet$ function call
				$\bullet$ method call
				$\bullet$ dynamic dispatch
				
				$\bullet$ polymorphism
				$\bullet$ closure
				& 
				$fun  ::= \cd{fun}~f~(\vec{x})\{e\} \mid\cd{call}(f,\vec{e})\mid\cd{pcall}(e',f,\vec{e})$ \\
				\hline
				$\bullet$ branch
				$\bullet$ pattern matching & 
				$branch::=\cd{case}\{\overrightarrow{e\rightarrow e'}\}$\\
				\hline
				$\bullet$ loop 
				$\bullet$ sequential & 
				$loop::=\cd{loop}~e~\{e'\}\mid\cd{break}\mid\cd{continue}$ \newline
				$seq::=e\cd{;} e'$\\
				\hline
				$\bullet$ concurrency
				& 
				$thread::=\cd{fork}(e)\mid \cd{wait}(e)$
				\\
				\hline
				$\bullet$ raw pointers $\bullet$ intrinsic function
				& $e::=mop$\\
				\hline
				$\bullet$ struct and enum types
				$\bullet$ generic types
				& supported by the translation\\
				\hline
				$\bullet$ Box
				$\bullet$ Array
				$\bullet$ Vector
				& $-$\\
				\bottomrule
		\end{tabular}}
	\end{center}
	\vspace{-0.5cm}
\end{table}

In CL, all constructs are expressions $e$ that can be evaluated to values.
The notion of
variables ($x$), values ($v$), dereferences ($*e$), fields ($e.e'$), arithmetic and boolean expressions,
print, skip (\cd{clskip}), and assertions are standard. 
%
Shared (\cd{\&}$e$) and mutable (\cd{\& mut}~$e$) references are the same as in Rust.

Functions in CL are defined as expressions that can be evaluated and assigned to variables.
This is designed to represent Rust closures,
which are anonymous functions with environments.
There are two forms of function calls: $\cd{call}$ and $\cd{pcall}$.
The former just calls the function $f$ with a sequence of arguments, whilst the latter is a polymorphic call $\cd{pcall}(e',f,\vec{e})$ implementing dynamic dispatchs.
For instance, in Rust there are expressions like $e'.f(e_1,\ldots,e_n)$, for which
we only know that $e'$ is an object implementing a trait (traits are like interfaces) containing $f$.
Since there may be more than one type implementing the trait,
the concrete function $f$ to be invoked depends on the type of $e'$ at runtime.
As we modeled the memory as typed, i.e. each memory location stores both values
and their types, the type of $e'$ at runtime can be obtained by reading the memory.
%
%
Branches in CL are defined by the \cd{case} construct.
Each case $e\rightarrow e'$ is a guarded action such that $e'$ can be executed only if $e$ evaluates to be true.
The constructs of threads ($thread$), bindings ($bindings$), assignment ($assign$), and sequence ($seq$) are standard.
The memory operations $mop$ are also constructs in CL.
It means that the memory access in CL invokes the memory operations in Section \ref{sec:mem-model}.
As the OBS is formalized in the memory model, CL semantics does not need to concern about it.

The translation semantics translates a Rust program to a CL program.
Table \ref{tab:correspondence-rust-cl} shows that
most features in Rust can be translated to corresponding CL constructs.
We explain the last two rows in the table.
For struct and enum types (user-defined types),
in CL, the definitions of user-defined types are not stored.
When a Rust program tries to create an object of a user-defined type,
the translation generates an allocation in memory block with size of the type in CL.
Translation semantics records the information of types, therefore
it knows the memory block size for each type. 
Three data structures: Box, Array, and Vector are not hard-coded in the semantics,
since they are just a collection of Rust programs.

Although RustSEM models a considerably large subset of Rust, it is not complete. Rust has no standard and evolves very fast, which makes it hard or almost impossible to some extent to build
a complete semantics for it. We do not model module definitions, \#[derive] attributes, among others.

Developing a formal semantics for real world languages always requires huge efforts. RustSEM is formalized in $\mathbb{K}$ and consists of around 1100  semantics rules, taking two and a half man-years.

\section{Evaluation of Semantics Correctness and Applications}
\label{sec:evaluation}

In this section, we evaluate the semantics correctness of RustSEM (Subsection \ref{subsec-rustsem-correctness}) and its applications in verification (Subsection \ref{subsec-rust-application}).

\subsection{Semantics Correctness}
\label{subsec-rustsem-correctness}



We have evaluated RustSEM in the following three aspects:
(1) \emph{Functional correctness.} For each input of a program, we compare
the execution output of RustSEM and the machine code generated by the Rust compiler.
(2) \emph{OBS checking evaluation.} For each program, we check whether RustSEM maintains the OBS invariants. It is compared with the compilation result of the Rust compiler.
(3) \emph{Semantic consistency}, i.e., the absence of semantic ambiguities.

We elaborate semantic ambiguities by an example.
Consider the \cd{if-construct}: ``\cd{if e B$_1$ else B$_2$}'' where \cd{e} is an expression and
$B_i$ is an execution block for $i=1,2$.
The  semantics for the construct could be:

\RuleEnvShort{Branch\label{Branch}}{
	$(C, e)\leadsto (C',v)$\\
	\hline
	$(C, \cd{if e }B_1\cd{ else }B_2)\leadsto (C', \cd{if v }B_1\cd{ else }B_2)$
	\quad\quad
}
\noindent
\TwoColumn{
	\ruleEnv{True\label{True}}{
		$\cd{v}=\emph{true}$\\
		\hline
		$(C, \cd{if v }B_1\cd{ else }B_2)\leadsto (C, B_1)$
		\quad\quad\quad\quad
	}
}{
	\ruleEnv{False\label{False}}{
		~\\
		\hline
		$(C, \cd{if v }B_1\cd{ else }B_2)\leadsto (C,B_2)$
	}
	
}

The semantics is a transition relation $(C,m)\leadsto (C',m')$, where $C,C'$ are configurations and $m,m'$ are terms, where a term could be a program or a value.
Rule \refrule{Branch} evaluates the expression $e$ to a value \cd{v} under the configuration $C$ 
and $C$ evolves into a new configuration $C'$ since the evaluation may have side effects. 
The \cd{if-construct} is reduced to another \cd{if-construct} with 
$e$ replaced by its value \cd{v}.
Rule \refrule{True} is for the case that \cd{v} equals \emph{true}, in which
the first block should be executed.
Rule \refrule{False} is for the case that \cd{v} equals \emph{false},
in which the second block should be executed.
Here we deliberately remove the checking $\cd{v}=\emph{false}$ from the premise of Rule \refrule{False},
which will lead to ambiguity.
%
Both Rule \refrule{True} and Rule \refrule{False} can match the \cd{if-construct} with $\cd{v}$ being $\emph{true}$.
This is called \emph{{semantic ambiguity}}, i.e., more than one semantics rules
could be applied to the same pair of configuration and term.
If $\mathbb{K}$ always tries to match Rule \refrule{True} first then the execution has no problem.
But if $\mathbb{K}$ matches Rule \refrule{False} first then the semantics is not correct.
This bug could be fixed by adding $\cd{v}=\emph{false}$ to the premises of \refrule{False}.

There is another {semantic ambiguity} in the example.
If we do not distinguish values from expressions in the semantics then
the construct ``\cd{if \emph{true} $B_1$ else $B_2$} '' can match both Rule \refrule{Branch}
and \refrule{True}, since \emph{true} is also an expression.
Rule \refrule{Branch} could match it infinitely many times, which should be avoided in our semantics.

Now we formally define the semantics correctness and explain our testing methods.

\begin{definition}[Semantics rules and semantics]
	Let $C$ be a set of configurations and $M$ be a set of terms.
	A \emph{semantics rule} is a partial function: $Ru:C\times M\xrightharpoonup{} C\times M$, i.e., mapping (or reducing) a pair $(c,m)$ of configuration $c$ and term $m$ to a new pair $(c',m')$ of configuration $c'$ and term $m'$.
	A \emph{semantics} is a set	of semantics rules: $SL=\{Ru_0,\ldots,Ru_n\}$.
\end{definition}

\begin{definition}[Execution traces of semantics]
	Let $SL$ be a semantics for a language.
	A finite execution trace of $SL$ is a sequence
	$\pi=(c_0,m_0)(c_1,m_1)\ldots(c_n,m_n)$, such that 
	for each $0\leq i < n$, $\exists Ru\in SL. Ru(c_i,m_i)=(c_{i+1},m_{i+1})$.
	The trace $\pi$ is called terminated if for all rule $Ru\in SL$, $Ru((c_n,m_n))$ is undefined.
	An infinite or divergence trace is an infinite sequence $\pi=(c_0,m_0)(c_1,m_1)\ldots$, such that 
	for each $0\leq i $, $\exists Ru\in SL. Ru(c_i,m_i)=(c_{i+1},m_{i+1})$.
\end{definition}

\begin{definition}[Semantics Correctness]
	Let $m$ be a program in a language, whose semantics is $SL$, the subset of configurations $pre_m\subseteq C$ be the precondition of $m$, the relation $post_m\subseteq C \times(C\cup \{\infty\})$ be the postcondition of $m$, where $\infty$ denotes that	the computation is diverge.
	The semantics $SL$ is called \emph{partially correct} iff
	for all program $m$ and 
	for all $c\in pre_m$, either there is a terminating trace $(c,m)\ldots(c',m')$ in $SL$ satisfying $((c,m),(c',m'))\in post_m$ or
	there is an infinite trace starting from $(c,m)$ satisfying $((c,m),\infty)\in post_m$.
	The semantics $SL$ is called \emph{completely correct} iff
	for all program $m$ and 
	for all $c\in pre_m$, every terminating traces $(c,m)\ldots(c',m')$ in $SL$ satisfying $((c,m),(c',m'))\in post_m$ and
	if there is an infinite trace starting from $(c,m)$ then $((c,m),\infty)\in post_m$.
\end{definition}

Let $SL$ be a semantics, $c$ be a configuration and $m$ be a term.
We denote by $\emph{Trace}_{SL}(c,m)$ the set of all terminating or infinite traces
with $(c,m)$ as the initial pair.
\kname's execution engine selects one of the traces in $\emph{Trace}_{SL}(c,m)$
for execution, whilst \kname's verification engine searches for all possible execution traces in $\emph{Trace}_{SL}(c,m)$.

An ambiguous semantics might be \emph{partially correct} but not \emph{completely correct}.
Thus, we need to detect {semantic ambiguities} to ensure \emph{complete correctness}.
Unlike Coq or Isabelle, $\mathbb{K}$ is not a theorem prover that can be used to prove some properties of RustSEM.
Instead, we can use its execution engine for testing. 
However,
only using its execution engine is not enough for detecting {semantic ambiguities}
if it always selects the right trace for execution.
Therefore, we also use its verification engine for testing.
Note that this is still called testing because test cases are used and the input of each test case is fixed. 
For each test case, the following three steps are carried out.
\begin{enumerate}
	\item Running the execution engine on the test case to detect bugs quickly.
	\item Running the verification engine to detect semantic ambiguities.
	\item Appending \cd{assert(False)} to the end of each test case and run the verification engine again.
\end{enumerate}
Note that Step (3) is used to detect semantic ambiguities missed by Steps (1) and (2). 
If a test case is not properly selected, it will be correct under all possible execution traces. Thus \kname~only outputs ``\cd{sat}'' and some {semantic ambiguities} are missed.
If we append \cd{assert(False)} at the end of the test case then the verification engine can report all execution traces to \cd{assert(False)} as the program always fails.

This testing strategy is efficient and effective. It is efficient because the test cases are small and the inputs are fixed, thus no state explosion exists.
It is effective because it can detect semantic ambiguities, even with very simple test cases. 
For instance, the following test case with the \cd{if-construct} could detect the two {semantic ambiguities} in Rules \refrule{Branch}, \refrule{True}, and \refrule{False} by running the verification engine.
\begin{verbatim}
               if (true)  { assert(true)  } else { assert(false) }
               if (false) { assert(false) } else { assert(true)  }
\end{verbatim}

We have used 400 test cases for the memory operations and CL semantics,
300 test cases for the translation semantics.
These test cases mainly come from the Rust benchmark \cite{benchmark}, the Rust libraries and the Rust text book \cite{rustbook}.
Besides, we have also created new test cases for testing our new grammars of the memory operations and CL.
The execution result of each test in RustSEM is compared with the execution result and compilation result of the Rust compiler version 1.4.5.
All experiments are conducted on a computer with an Intel Xeon(R) CPU E5-1650 v3 @ 3.50GHz $\times$ 12 and 16GB DDR4 RAM.
The
semantics of one construct in Rust is usually formalized with a collection of $\mathbb{K}$ rules, and thus one test case can cover a collection of rules. 

The experimental results show that:
\begin{enumerate}
	\item Some correct tests are accepted by RustSEM but rejected by the Rust compiler, due to the conservative analysis of the compiler.
	\item Some incorrect tests are rejected by RustSEM but accepted by the Rust compiler, showing that RustSEM is more powerful in detecting memory errors, especially in the programs mixed with safe and unsafe operations. For example, after the ownership of a block has been moved away from a variable by a safe operation, a raw pointer reads the block through the reference to the variable. 
	\item More than 36 {semantic ambiguities} are detected, showing that our testing technique for ambiguities is effective.
	\item For a collection of tests with memory errors, RustSEM can successfully detect all errors. More details are discussed in the application in runtime verification (Subsection \ref{subsec-rust-application}).
\end{enumerate}


\subsection{Applications in Verification}
\label{subsec-rust-application}

\emph{Runtime Verification for Detecting Memory Errors.} 
RustSEM provides a runtime checker for detecting memory errors, since the memory model integrates checkings for invalid memory accesses as explained in Section~\ref{sec:mem-model}. It is  based on semantics instrumentation rather than most of the existing runtime checkers, such as \cite{DBLP:conf/issta/ChenYKQX19}, which are based on code instrumentation. 
RustSEM detects: (1)
accessing uninitialized locations, (2) reading dangling pointers, (3) double frees, (4) data races and deadlocks, (5) ownership and borrowing errors, (6) buffer overflows. 

\begin{wraptable}{r}{8.2cm}
	\caption{Results for runtime verification.}\label{tab-runtime}
	{\small
		\begin{tabular}{p{3cm}|p{0.5cm}|p{0.5cm}|p{1cm}|p{1.1cm}}
			\hline
			\emph{Memory errors} & TC & IN & \emph{Avg. Time$(s)$} &  \emph{Avg. Mem$($KB$)$}  \\
			\hline\hline
			read
			uninit. values & 22  & 10 & 6.16 & 387240\\
			\hline
			read dangling pointers & 11 & 10 & 5.89 & 390356\\
			\hline
			double frees & 10  & 10 & 5.73 & 346096 \\
			\hline
			data races & 14  & 10 &7.36 & 394152\\
			\hline
			ownership \& borrowing & 40 & 10 & 6.64 & 344088 \\
			\hline
			index overflows & 21  & 10 &7.16 & 379788\\
			\hline
		\end{tabular}
	}
\end{wraptable}
Table \ref{tab-runtime} shows the evaluation result for runtime verification.
``TC'', ``IN'', ``Avg. Time(s)'', and ``Avg. Mem(KB)'' denote the number of test cases, the number of inputs for each test case, average execution time and average memory consumption, respectively.
For each kind of detectable memory errors, a collection of test cases is created.
All the memory errors in the test cases can be detected with the inputs that can trigger the errors.
Data races are not always detected in an execution, since executions are non-deterministic.


Compared with other runtime checkers, RustSEM is less efficient and depends on the efficiency of \kname~execution engine.
In order to execute a program , \kname~first constructs its abstract syntax tree and searches the corresponding semantics rules for the tree at each execution step. 
The advantage of RustSEM is that the memory checking is based on semantics instrumentation
instead of code instrumentation. 
The biggest test case is around 210 LOC with the execution time of  around 10 seconds, including both
translation and execution time.

\emph{Formal Verification for both Memory and Functional Properties.}
\kname~has a verification engine, which can be used to construct a program verifier for a language by instantiating it with an operational semantics of that language.
The basic idea of the infrastructure could be explained as follows.
\kname~verification engine takes an operational
semantics given in \kname~and generates queries to a theorem
prover (for example, Z3 \cite{DBLP:conf/tacas/MouraB08}). The program correctness properties are given as reachability rules between matching 
patterns in \kname~semantics rules. Internally, the verifier uses the operational semantics to perform symbolic execution. Also, it has an internal matching logic prover for reasoning about implication
between patterns (states), which reduces to SMT reasoning (refer to \kname~verification infrastructure \cite{oopsla/StefanescuPYLR16} for more details).


For the verification, it is necessary to specify a pre- and post-condition similar to Hoare logic, and loops need to be annotated with  invariants.
For simplicity, we show the verified properties as a pair $\{Precondition\}\{Postcondition\}$, indicating that if the input of the program satisfies the precondition then the result satisfies the postcondition.
Thanks to the memory access checking formalized in RustSEM,
we can uniformly {specify} a memory property, such as data races, buffer overflows, or borrowing errors,
as \{$True$\}\{$\neg stuck$\},
which means that for any input, the program cannot get stuck.
For other kinds of properties, such as functional properties,
the program should satisfy both the memory guarantees of RustSEM
and the properties being specified.
\begin{wraptable}{r}{7.6cm}
	\centering
	\caption{Results for formal verification}
	\label{tab:verification-results}
	{\small
		\begin{tabular}{c|l|c|c|c}
			\hline
			\multicolumn{2}{c|}{Programs} &  ~LOC~ & ~Time(s)~ & ~Mem(KB)~\\
			\hline\hline 
			\multicolumn{2}{c|}{\cd{sum($N$)}} & 15  & 44.65  & 698240 \\ 
			\hline
			\multicolumn{2}{c|}{\cd{sumvec($V$)}}   & 18  & 63.92 & 1470880 \\
			\hline
			\multicolumn{2}{c|}{\cd{insertion\_sort($V$)}} & 43 & 125.43 & 8736592\\
			\hline
			\multicolumn{2}{c|}{polymorphism} & 60 & 42.20 & 678992 \\
			\hline
			\multicolumn{2}{c|}{closure} & 22 & 36.45 & 663268 \\
			\hline
			\multicolumn{2}{c|}{trait} & 50 & 39.22 & 654672 \\
			\hline
			\multicolumn{2}{c|}{concur-race} & 58 & 47.25 & 647328 \\
			\hline
			\multicolumn{2}{c|}{concur-order} &  57 &  60.82 & 2931760 \\
			\hline
			\multirow{4}{*}{VecDeque} & new & \multirow{4}{*}{210} & 76.90 & 1160640 \\
			\cline{2-2}
			\cline{4-5}
			& push\_front & & 115.33 & 1667140 \\
			\cline{2-2}
			\cline{4-5}
			& push\_back & & 140.70 & 2025244 \\
			\cline{2-2}
			\cline{4-5}
			& reserve & & 85 & 1859304 \\
			\hline
	\end{tabular}}
	\vspace{-0.5cm}
\end{wraptable}

Table \ref{tab:verification-results} shows the verification results of a number of programs verified in RustSEM using the $\mathbb{K}$ verifier, where
LOC, Time(s) and Mem(KB) denote the lines of code, execution time, and 
memory consumption, respectively. 
``\cd{sum($N$)}" is a function that computes the sum from $1$ to the input $N$, verifying $\{N\geq 1\}\{N\times(N+1)/2\}$.
``\cd{sumvec($V$)}" computes the sum of all elements in $V$, and we verify $\{[v_1,\ldots,v_n]\in\cd{vec<i32>}\}\{\Sigma_{1\leq i\leq n} v_i\}$, with the precondition being a vector of $n$ random integers, and the postcondition the sum of its elements.
``\cd{insertion\_sort}" implements an {insertion sort algorithm}, where we verify $\{V\in\cd{vec<i32>}\}\{V'\in\cd{vec<i32>}~s.t.~\cd{sameElem}(V,V')\land \cd{order}(V')\}$. The precondition is an integer vector and the postcondition is new vector with the same elements (checked by \cd{sameElem}) and the vector is ordered (checked by \cd{order}).
The program ``\cd{concur-race}" is a multi-threaded program with a data race bug where we verify $\{True\}\{\neg stuck\}$, that is the program finishes without going into a stuck state.
The program has a data race, and the verification can successfully detect it, going to a stuck state thanks to the invalid access checking in the memory model, hence the property is invalid.
The program ``\cd{concur-order}" is a multi-thread program without data races.
The property is the same as for ``\cd{concur-race}".
The verifier explores all possible interleavings in the concurrent programs to detect bugs.
\emph{In RustSEM, we optimized the semantics to allow interleaving only if the rules access some shared resources.}

The program ``VecDeque" is a module from the Rust standard library, for which a previous version~\cite{VECDEQUE} contained a bug. We verify this old version implementation of VecDeque to rediscover the error by verifying the functions \cd{new}, \cd{push\_front}, \cd{push\_back}, and \cd{reserve}.
The property is: $\{0\leq \cd{head} < \cd{cap}\land 0\leq \cd{tail} < \cd{cap}\} \{0\leq \cd{head$'$} < \cd{cap$'$}\land 0\leq \cd{tail$'$} <\cd{cap$'$}\}$.
VecDeque has a \cd{head} to indicate the memory location for \cd{push\_front} and 
and a \cd{tail} to indicate the memory location for \cd{push\_back}.
The variable \cd{cap} is the capacity of the VecDeque.
The property requires that the \cd{head} and \cd{tail} of a VecDeque have to
be within the range $[0,\cd{cap})$ before and after executing any of its methods.
The primed variables denote the variables' values after executing the methods.
The bug \cite{VECDEQUE} can be rediscovered by RustSEM in the method \cd{reserve}  resulting
in a buffer overflow.
RustSEM does not support infinite heap structure specification and verification now.
For the verification we use an approach similar to bounded model checking, setting a bound in the maximum capacity of 
\cd{vecDeque} to 16, limiting the state space exploration.
The benchmarks \cd{polymorphism}, \cd{closure}, and \cd{trait} are programs using polymorphism, closures, and traits, respectively and we verify some functional properties.


Now, we discuss the limitations of formal verification.
Firstly, the property specification language in \kname~needs to specify pre- and post-conditions and loop invariants for each memory locations, which needs a lot of efforts.
We only give an abstract description of the properties being verified here.
More details of specification languages could be found in
\kname~verification infrastructure \cite{oopsla/StefanescuPYLR16}.
Secondly, we cannot specify infinite heap data structures, such as trees.
\kname~has the capability to reason about infinite heap data structures, such as binary search trees,
which is shown in \kname~verification infrastructure \cite{oopsla/StefanescuPYLR16}.
The challenge is that \kname~verification infrastructure uses a memory model different from RustSEM, as RustSEM incorporates OBS.
Therefore, we cannot directly inherit the framework from it.
Actually, more work needs to be done to define a language that could
be used to inductively define infinite data structures in RustSEM.
One solution is to
discover the relations
between RustSEM's memory model and separation logic \cite{DBLP:conf/popl/IshtiaqO01}\cite{DBLP:conf/lics/Reynolds02}.
In this paper, we focus on the semantics of Rust.
A more user-friendly and powerful verifier is our future work.

\section{Related Work}
\label{sec:relatedwork}

\paragraph{\bf Rust semantics}

Table \ref{tab-relatedwork1} compares the existing semantics for Rust on their target languages (lang), the support of major features including safe and unsafe constructs, concurrency (concur), dynamic OBS (D-OBS) and type system, as well as their verification capabilities, i.e., whether support automated verification (AV) and machine-checked manual verification (MV).

RustSEM and KRust are the only two that directly formalize the Rust language, whereas all the remaining semantics, i.e., RustBelt, Oxide, Patina and K-Rust, formalizes either a variant or an outdated version of Rust.

RustSEM covers more major features than exiting semantics.
Indeed, RustSEM fully supports both safe and unsafe constructs and concurrency. In contrast, existing semantics only partially supports them. For instance, RustBelt does not formalize some advanced features such as trait objects, as it is for a variant of Rust, while Oxide, Patina, K-Rust and KRust only support a subset of safe constructs without concurrency.
Furthermore, RustSEM is the only one supporting dynamic OBS.
RustSEM focuses the operational semantics of OBS, thus the type system is not included.

RustSEM, K-Rust and KRust can be used for automated verification.
Although RustBelt, implemented in the Coq theorem prover, can verify $\lambda_{Rust}$ programs using Coq, the verification is interactive and needs expertise and manual inspection.
Patina and Oxide have not yet been implemented in any tool. Thus they provide neither automated nor machine-checked verification.

\paragraph{\bf Rust verification}
CRust~\cite{TomanPT15}, SMACK-Rust~\cite{DBLP:conf/atva/BaranowskiHR18} and Viper-based Prusti \cite{AstrauskasMuellerPoliSummers19b} can verify Rust programs by translating them into the input languages of existing verification tools, instead of building formal semantics for Rust. 
For instance,
SMACK-Rust compiles Rust programs into the LLVM code which can be verified by SMACK \cite{DBLP:conf/icse/CarterHWRE16}. 
Prusti translates a subset of safe constructs to the intermediate language of Viper  \cite{mcai/0001SS16} to construct core proofs.
Note that these tools cover less features than RustSEM.

Stackborrow \cite{Ralf-Jung2020} presents an alias model to regulate the use of unsafe pointers, and checks borrowings by creating a stack for each memory location, instead of explicitly
using lifetimes.
Moreover, Stackborrow does not support concurrency.
Stackborrow is implemented in Miri \cite{MIRI}, which is not a formal implementation.

{
	\paragraph{\bf Ownership and Borrowing}
	The concept of ownership has been proposed for many years and 
	most of related works exploit type systems to enforce various ownership disciplines.
	Cyclone \cite{DBLP:conf/usenix/JimMGHCW02} is designed to be a safe dialect of the C language by using region based memory management \cite{DBLP:conf/popl/TofteT94}.
	Mezzo \cite{DBLP:journals/toplas/BalabonskiPP16} and Alms \cite{DBLP:conf/popl/TovP111} follow ML-tradition and employ substructural type systems for managing ownership.
	There are also many type systems associating reading and writing permissions to aliases \cite{DBLP:conf/ecoop/BoylandNR011}, such as 
	Mezzo \cite{DBLP:journals/toplas/BalabonskiPP16}, Pony \cite{DBLP:conf/agere/ClebschDBM15} and AEminium \cite{DBLP:conf/pldi/StorkNSMFMA14}.
	All these works have different technique details with Rust and exploit types systems to enforce memory safety.
	Our work aims at explaining the techniques of Rust from an operational aspects instead of type systems
	in order to provide a new insight of OBS.}

\newcommand{\psupport}{
	\Large
	\RIGHTcircle
}
\newcommand{\fsupport}{
	\Large
	\CIRCLE
}
{\small
	\begin{table}[t]
		\caption{The comparison with related semantics works}
		\label{tab-relatedwork1}
		{
			\begin{tabular}{p{4cm}|ccccccccc}
				\hline
				& \multirow{2}{*}{lang} & \multicolumn{5}{c}{features} && \multicolumn{2}{c}{Verification} \\
				\cline{3-7}
				\cline{9-10}
				&  & safe & unsafe & type & D-OBS & concur & & AV & MV\\
				\hline
				RustBelt \cite{Ralf2018}  & $\lambda_{Rust}$ & \psupport &\psupport & \fsupport &   &  \fsupport  &  &  & \Checkmark \\
				\hline
				Patina  \cite{Eric2015} & old Rust & \psupport &   & \psupport  \\
				\hline
				Oxide \cite{DBLP:journals/corr/abs-1903-00982} & Oxide& \psupport  & & \psupport  &&  \\
				\hline
				KRust \cite{corr/abs-1804-10806} & Rust & \psupport & & & &  & & \Checkmark  \\
				\hline
				K-Rust\cite{DBLP:journals/corr/abs-1804-07608} & core-lang & \psupport &  
				& \psupport & & & &  \Checkmark & \\
				\hline
				RustSEM & Rust & \fsupport &  \fsupport & & \fsupport & \fsupport  & & \Checkmark  &  \\
				\hline
			\end{tabular}\\
			{\psupport}~partially supported\quad\quad
			{\fsupport}~fully supported}
		\vspace{-0.5cm}
\end{table}}

\section{Conclusion}
\label{sec:conclusion}

We have proposed a high-level abstraction of OBS and an executable operational semantics for Rust.
Compared with existing works, it covers a larger subset of the major language features of Rust.
The core of the semantics is the memory model with the operational semantics of OBS, which can be potentially reused by other languages to improve memory safety.
We have proved the refinement relation between the high-level abstraction of OBS and the operational semantics in the memory model.
Moreover, we proposed a technique for testing semantic consistency in order to detect semantic ambiguities and provide completely correct semantics.
We have shown that RustSEM can be applied in automated runtime and formal verification.

As future works, 
we are working on two directions.
The first one is to
extend the verification infrastructure of \kname~to support infinite heap structure verification for the memory model of RustSEM.
The second one is to
reuse our memory model for the \kname~semantics of the C programming language to improve C memory safety by checking the preservation of the OBS invariants.

\bibliography{reference}


\begin{thebibliography}{38}


\ifx \showCODEN    \undefined \def \showCODEN     #1{\unskip}     \fi
\ifx \showDOI      \undefined \def \showDOI       #1{#1}\fi
\ifx \showISBNx    \undefined \def \showISBNx     #1{\unskip}     \fi
\ifx \showISBNxiii \undefined \def \showISBNxiii  #1{\unskip}     \fi
\ifx \showISSN     \undefined \def \showISSN      #1{\unskip}     \fi
\ifx \showLCCN     \undefined \def \showLCCN      #1{\unskip}     \fi
\ifx \shownote     \undefined \def \shownote      #1{#1}          \fi
\ifx \showarticletitle \undefined \def \showarticletitle #1{#1}   \fi
\ifx \showURL      \undefined \def \showURL       {\relax}        \fi
\providecommand\bibfield[2]{#2}
\providecommand\bibinfo[2]{#2}
\providecommand\natexlab[1]{#1}
\providecommand\showeprint[2][]{arXiv:#2}

\bibitem[\protect\citeauthoryear{Astrauskas, M\"uller, Poli, and
  Summers}{Astrauskas et~al\mbox{.}}{2019}]%
        {AstrauskasMuellerPoliSummers19b}
\bibfield{author}{\bibinfo{person}{V. Astrauskas}, \bibinfo{person}{P.
  M\"uller}, \bibinfo{person}{F. Poli}, {and} \bibinfo{person}{A.~J. Summers}.}
  \bibinfo{year}{2019}\natexlab{}.
\newblock \showarticletitle{{Leveraging {R}ust Types for Modular Specification
  and Verification}}. In \bibinfo{booktitle}{\emph{to appear in Object-Oriented
  Programming Systems, Languages, and Applications (OOPSLA)}}.
  \bibinfo{publisher}{ACM}.
\newblock
\urldef\tempurl%
\url{https://doi.org/10.1145/3360573}
\showDOI{\tempurl}


\bibitem[\protect\citeauthoryear{Balabonski, Pottier, and Protzenko}{Balabonski
  et~al\mbox{.}}{2016}]%
        {DBLP:journals/toplas/BalabonskiPP16}
\bibfield{author}{\bibinfo{person}{Thibaut Balabonski},
  \bibinfo{person}{Fran{\c{c}}ois Pottier}, {and} \bibinfo{person}{Jonathan
  Protzenko}.} \bibinfo{year}{2016}\natexlab{}.
\newblock \showarticletitle{{The Design and Formalization of Mezzo, a
  Permission-Based Programming Language}}.
\newblock \bibinfo{journal}{\emph{{ACM} Trans. Program. Lang. Syst.}}
  \bibinfo{volume}{38}, \bibinfo{number}{4} (\bibinfo{year}{2016}),
  \bibinfo{pages}{14:1--14:94}.
\newblock
\urldef\tempurl%
\url{http://dl.acm.org/citation.cfm?id=2837022}
\showURL{%
\tempurl}


\bibitem[\protect\citeauthoryear{Baranowski, He, and Rakamaric}{Baranowski
  et~al\mbox{.}}{2018}]%
        {DBLP:conf/atva/BaranowskiHR18}
\bibfield{author}{\bibinfo{person}{Marek~S. Baranowski},
  \bibinfo{person}{Shaobo He}, {and} \bibinfo{person}{Zvonimir Rakamaric}.}
  \bibinfo{year}{2018}\natexlab{}.
\newblock \showarticletitle{{Verifying Rust Programs with {SMACK}}}. In
  \bibinfo{booktitle}{\emph{Automated Technology for Verification and Analysis
  - 16th International Symposium, {ATVA} 2018, Los Angeles, CA, USA, October
  7-10, 2018, Proceedings}} \emph{(\bibinfo{series}{Lecture Notes in Computer
  Science})}, \bibfield{editor}{\bibinfo{person}{Shuvendu~K. Lahiri} {and}
  \bibinfo{person}{Chao Wang}} (Eds.), Vol.~\bibinfo{volume}{11138}.
  \bibinfo{publisher}{Springer}, \bibinfo{pages}{528--535}.
\newblock
\showISBNx{978-3-030-01089-8}
\urldef\tempurl%
\url{https://doi.org/10.1007/978-3-030-01090-4\_32}
\showDOI{\tempurl}


\bibitem[\protect\citeauthoryear{Bogd\u{a}na\c{s} and
  Ro\c{s}u}{Bogd\u{a}na\c{s} and Ro\c{s}u}{2015}]%
        {bogdanas-rosu-2015-popl}
\bibfield{author}{\bibinfo{person}{Denis Bogd\u{a}na\c{s}} {and}
  \bibinfo{person}{Grigore Ro\c{s}u}.} \bibinfo{year}{2015}\natexlab{}.
\newblock \showarticletitle{{{K-Java: A Complete Semantics of Java}}}. In
  \bibinfo{booktitle}{\emph{Proceedings of the 42nd Symposium on Principles of
  Programming Languages (POPL'15)}}. \bibinfo{publisher}{ACM},
  \bibinfo{pages}{445--456}.
\newblock
\urldef\tempurl%
\url{https://doi.org/10.1145/2676726.2676982}
\showDOI{\tempurl}


\bibitem[\protect\citeauthoryear{Boyland, Noble, and Retert}{Boyland
  et~al\mbox{.}}{2001}]%
        {DBLP:conf/ecoop/BoylandNR011}
\bibfield{author}{\bibinfo{person}{John Boyland}, \bibinfo{person}{James
  Noble}, {and} \bibinfo{person}{William Retert}.}
  \bibinfo{year}{2001}\natexlab{}.
\newblock \showarticletitle{Capabilities for Sharing: {A} Generalisation of
  Uniqueness and Read-Only}. In \bibinfo{booktitle}{\emph{{ECOOP} 2001 -
  Object-Oriented Programming, 15th European Conference, Budapest, Hungary,
  June 18-22, 2001, Proceedings}} \emph{(\bibinfo{series}{Lecture Notes in
  Computer Science})}, \bibfield{editor}{\bibinfo{person}{J{\o}rgen~Lindskov
  Knudsen}} (Ed.), Vol.~\bibinfo{volume}{2072}. \bibinfo{publisher}{Springer},
  \bibinfo{pages}{2--27}.
\newblock
\urldef\tempurl%
\url{https://doi.org/10.1007/3-540-45337-7\_2}
\showDOI{\tempurl}


\bibitem[\protect\citeauthoryear{Carter, He, Whitaker, Rakamaric, and
  Emmi}{Carter et~al\mbox{.}}{2016}]%
        {DBLP:conf/icse/CarterHWRE16}
\bibfield{author}{\bibinfo{person}{Montgomery Carter}, \bibinfo{person}{Shaobo
  He}, \bibinfo{person}{Jonathan Whitaker}, \bibinfo{person}{Zvonimir
  Rakamaric}, {and} \bibinfo{person}{Michael Emmi}.}
  \bibinfo{year}{2016}\natexlab{}.
\newblock \showarticletitle{{{SMACK} software verification toolchain}}. In
  \bibinfo{booktitle}{\emph{Proceedings of the 38th International Conference on
  Software Engineering, {ICSE} 2016, Austin, TX, USA, May 14-22, 2016 -
  Companion Volume}}, \bibfield{editor}{\bibinfo{person}{Laura~K. Dillon},
  \bibinfo{person}{Willem Visser}, {and} \bibinfo{person}{Laurie Williams}}
  (Eds.). \bibinfo{publisher}{{ACM}}, \bibinfo{pages}{589--592}.
\newblock
\showISBNx{978-1-4503-4205-6}
\urldef\tempurl%
\url{https://doi.org/10.1145/2889160.2889163}
\showDOI{\tempurl}


\bibitem[\protect\citeauthoryear{Chen, Yan, Kan, Qian, and Xue}{Chen
  et~al\mbox{.}}{2019}]%
        {DBLP:conf/issta/ChenYKQX19}
\bibfield{author}{\bibinfo{person}{Zhe Chen}, \bibinfo{person}{Junqi Yan},
  \bibinfo{person}{Shuanglong Kan}, \bibinfo{person}{Ju Qian}, {and}
  \bibinfo{person}{Jingling Xue}.} \bibinfo{year}{2019}\natexlab{}.
\newblock \showarticletitle{{Detecting memory errors at runtime with
  source-level instrumentation}}. In \bibinfo{booktitle}{\emph{Proceedings of
  the 28th {ACM} {SIGSOFT} International Symposium on Software Testing and
  Analysis, {ISSTA} 2019, Beijing, China, July 15-19, 2019.}},
  \bibfield{editor}{\bibinfo{person}{Dongmei Zhang} {and}
  \bibinfo{person}{Anders M{\o}ller}} (Eds.). \bibinfo{publisher}{{ACM}},
  \bibinfo{pages}{341--351}.
\newblock
\showISBNx{978-1-4503-6224-5}
\urldef\tempurl%
\url{https://doi.org/10.1145/3293882.3330581}
\showDOI{\tempurl}


\bibitem[\protect\citeauthoryear{Clebsch, Drossopoulou, Blessing, and
  McNeil}{Clebsch et~al\mbox{.}}{2015}]%
        {DBLP:conf/agere/ClebschDBM15}
\bibfield{author}{\bibinfo{person}{Sylvan Clebsch}, \bibinfo{person}{Sophia
  Drossopoulou}, \bibinfo{person}{Sebastian Blessing}, {and}
  \bibinfo{person}{Andy McNeil}.} \bibinfo{year}{2015}\natexlab{}.
\newblock \showarticletitle{{Deny capabilities for safe, fast actors}}. In
  \bibinfo{booktitle}{\emph{Proceedings of the 5th International Workshop on
  Programming Based on Actors, Agents, and Decentralized Control, AGERE! 2015,
  Pittsburgh, PA, USA, October 26, 2015}},
  \bibfield{editor}{\bibinfo{person}{Elisa~Gonzalez Boix},
  \bibinfo{person}{Philipp Haller}, \bibinfo{person}{Alessandro Ricci}, {and}
  \bibinfo{person}{Carlos Varela}} (Eds.). \bibinfo{publisher}{{ACM}},
  \bibinfo{pages}{1--12}.
\newblock
\urldef\tempurl%
\url{https://doi.org/10.1145/2824815.2824816}
\showDOI{\tempurl}


\bibitem[\protect\citeauthoryear{\c{S}tef\u{a}nescu, Park, Yuwen, Li, and
  Ro\c{s}u}{\c{S}tef\u{a}nescu et~al\mbox{.}}{2016}]%
        {oopsla/StefanescuPYLR16}
\bibfield{author}{\bibinfo{person}{Andrei \c{S}tef\u{a}nescu},
  \bibinfo{person}{Daejun Park}, \bibinfo{person}{Shijiao Yuwen},
  \bibinfo{person}{Yilong Li}, {and} \bibinfo{person}{Grigore Ro\c{s}u}.}
  \bibinfo{year}{2016}\natexlab{}.
\newblock \showarticletitle{{Semantics-Based Program Verifiers for All
  Languages}}. In \bibinfo{booktitle}{\emph{Proceedings of the 31th Conference
  on Object-Oriented Programming, Systems, Languages, and Applications
  (OOPSLA'16)}}. \bibinfo{publisher}{ACM}, \bibinfo{pages}{74--91}.
\newblock
\urldef\tempurl%
\url{https://doi.org/10.1145/2983990.2984027}
\showDOI{\tempurl}


\bibitem[\protect\citeauthoryear{Davidoff}{Davidoff}{2018}]%
        {Rustlibrarybug}
\bibfield{author}{\bibinfo{person}{Sergey~"Shnatsel" Davidoff}.}
  \bibinfo{year}{2018}\natexlab{}.
\newblock \bibinfo{title}{{How Rust’s standard library was vulnerable for
  years and nobody noticed}}.
\newblock
  \bibinfo{howpublished}{\url{https://medium.com/@shnatsel/how-rusts-standard-library-was-vulnerable-for-years-and-nobody-noticed-aebf0503c3d6}}.
\newblock


\bibitem[\protect\citeauthoryear{de~Moura and Bj{\o}rner}{de~Moura and
  Bj{\o}rner}{2008}]%
        {DBLP:conf/tacas/MouraB08}
\bibfield{author}{\bibinfo{person}{Leonardo~Mendon{\c{c}}a de Moura} {and}
  \bibinfo{person}{Nikolaj Bj{\o}rner}.} \bibinfo{year}{2008}\natexlab{}.
\newblock \showarticletitle{{Z3:} An Efficient {SMT} Solver}. In
  \bibinfo{booktitle}{\emph{Tools and Algorithms for the Construction and
  Analysis of Systems, 14th International Conference, {TACAS} 2008, Held as
  Part of the Joint European Conferences on Theory and Practice of Software,
  {ETAPS} 2008, Budapest, Hungary, March 29-April 6, 2008. Proceedings}}
  \emph{(\bibinfo{series}{Lecture Notes in Computer Science})},
  \bibfield{editor}{\bibinfo{person}{C.~R. Ramakrishnan} {and}
  \bibinfo{person}{Jakob Rehof}} (Eds.), Vol.~\bibinfo{volume}{4963}.
  \bibinfo{publisher}{Springer}, \bibinfo{pages}{337--340}.
\newblock
\showISBNx{978-3-540-78799-0}
\urldef\tempurl%
\url{https://doi.org/10.1007/978-3-540-78800-3\_24}
\showDOI{\tempurl}


\bibitem[\protect\citeauthoryear{Ellison and Rosu}{Ellison and Rosu}{2012}]%
        {ellison-rosu-2012-popl}
\bibfield{author}{\bibinfo{person}{Chucky Ellison} {and}
  \bibinfo{person}{Grigore Rosu}.} \bibinfo{year}{2012}\natexlab{}.
\newblock \showarticletitle{An Executable Formal Semantics of C with
  Applications}. In \bibinfo{booktitle}{\emph{Proceedings of the 39th ACM
  SIGPLAN-SIGACT Symposium on Principles of Programming Languages (POPL'12)}}.
  \bibinfo{publisher}{ACM}, \bibinfo{pages}{533--544}.
\newblock
\urldef\tempurl%
\url{https://doi.org/10.1145/2103656.2103719}
\showDOI{\tempurl}


\bibitem[\protect\citeauthoryear{Hathhorn, Ellison, and Ro\c{s}u}{Hathhorn
  et~al\mbox{.}}{2015}]%
        {hathhorn-ellison-rosu-2015-pldi}
\bibfield{author}{\bibinfo{person}{Chris Hathhorn}, \bibinfo{person}{Chucky
  Ellison}, {and} \bibinfo{person}{Grigore Ro\c{s}u}.}
  \bibinfo{year}{2015}\natexlab{}.
\newblock \showarticletitle{Defining the Undefinedness of C}. In
  \bibinfo{booktitle}{\emph{Proceedings of the 36th ACM SIGPLAN Conference on
  Programming Language Design and Implementation (PLDI'15)}}.
  \bibinfo{publisher}{ACM}, \bibinfo{pages}{336--345}.
\newblock
\urldef\tempurl%
\url{https://doi.org/10.1145/2813885.2737979}
\showDOI{\tempurl}


\bibitem[\protect\citeauthoryear{Ishtiaq and O'Hearn}{Ishtiaq and
  O'Hearn}{2001}]%
        {DBLP:conf/popl/IshtiaqO01}
\bibfield{author}{\bibinfo{person}{Samin~S. Ishtiaq} {and}
  \bibinfo{person}{Peter~W. O'Hearn}.} \bibinfo{year}{2001}\natexlab{}.
\newblock \showarticletitle{{{BI} as an Assertion Language for Mutable Data
  Structures}}. In \bibinfo{booktitle}{\emph{Conference Record of {POPL} 2001:
  The 28th {ACM} {SIGPLAN-SIGACT} Symposium on Principles of Programming
  Languages, London, UK, January 17-19, 2001}},
  \bibfield{editor}{\bibinfo{person}{Chris Hankin} {and} \bibinfo{person}{Dave
  Schmidt}} (Eds.). \bibinfo{publisher}{{ACM}}, \bibinfo{pages}{14--26}.
\newblock
\showISBNx{1-58113-336-7}
\urldef\tempurl%
\url{http://dl.acm.org/citation.cfm?id=360204}
\showURL{%
\tempurl}


\bibitem[\protect\citeauthoryear{issue 44800}{issue 44800}{2017}]%
        {VECDEQUE}
\bibfield{author}{\bibinfo{person}{VecDeque issue 44800}.}
  \bibinfo{year}{2017}\natexlab{}.
\newblock
  \bibinfo{howpublished}{\url{https://github.com/rust-lang/rust/issues/44800}}.
\newblock


\bibitem[\protect\citeauthoryear{Jim, Morrisett, Grossman, Hicks, Cheney, and
  Wang}{Jim et~al\mbox{.}}{2002}]%
        {DBLP:conf/usenix/JimMGHCW02}
\bibfield{author}{\bibinfo{person}{Trevor Jim}, \bibinfo{person}{J.~Gregory
  Morrisett}, \bibinfo{person}{Dan Grossman}, \bibinfo{person}{Michael~W.
  Hicks}, \bibinfo{person}{James Cheney}, {and} \bibinfo{person}{Yanling
  Wang}.} \bibinfo{year}{2002}\natexlab{}.
\newblock \showarticletitle{{Cyclone: {A} Safe Dialect of {C}}}. In
  \bibinfo{booktitle}{\emph{Proceedings of the General Track: 2002 {USENIX}
  Annual Technical Conference, June 10-15, 2002, Monterey, California, {USA}}},
  \bibfield{editor}{\bibinfo{person}{Carla~Schlatter Ellis}} (Ed.).
  \bibinfo{publisher}{{USENIX}}, \bibinfo{pages}{275--288}.
\newblock
\urldef\tempurl%
\url{http://www.usenix.org/publications/library/proceedings/usenix02/jim.html}
\showURL{%
\tempurl}


\bibitem[\protect\citeauthoryear{Jung, Dang, Kang, and Dreyer}{Jung
  et~al\mbox{.}}{2020}]%
        {Ralf-Jung2020}
\bibfield{author}{\bibinfo{person}{Ralf Jung}, \bibinfo{person}{Hoang-Hai
  Dang}, \bibinfo{person}{Jeehoon Kang}, {and} \bibinfo{person}{Derek Dreyer}.}
  \bibinfo{year}{2020}\natexlab{}.
\newblock \showarticletitle{{Stacked Borrows: An Aliasing Model for Rust}}. In
  \bibinfo{booktitle}{\emph{to appear in POPL 2020}}. \bibinfo{publisher}{ACM}.
\newblock


\bibitem[\protect\citeauthoryear{Jung, Jourdan, Krebbers, and Dreyer}{Jung
  et~al\mbox{.}}{2018a}]%
        {Ralf2018}
\bibfield{author}{\bibinfo{person}{Ralf Jung}, \bibinfo{person}{Jacques-Henri
  Jourdan}, \bibinfo{person}{Robbert Krebbers}, {and} \bibinfo{person}{Derek
  Dreyer}.} \bibinfo{year}{2018}\natexlab{a}.
\newblock \showarticletitle{{RustBelt: Securing the Foundations of the Rust
  Programming Language}}.
\newblock \bibinfo{journal}{\emph{{Proc. ACM Program. Lang. 2, POPL, Article}}}
  (\bibinfo{date}{Jan.} \bibinfo{year}{2018}).
\newblock
\urldef\tempurl%
\url{https://doi.org/10.1145/3158154}
\showDOI{\tempurl}


\bibitem[\protect\citeauthoryear{Jung, Krebbers, Jourdan, Bizjak, Birkedal, and
  Dreyer}{Jung et~al\mbox{.}}{2018b}]%
        {DBLP:journals/jfp/JungKJBBD18}
\bibfield{author}{\bibinfo{person}{Ralf Jung}, \bibinfo{person}{Robbert
  Krebbers}, \bibinfo{person}{Jacques{-}Henri Jourdan}, \bibinfo{person}{Ales
  Bizjak}, \bibinfo{person}{Lars Birkedal}, {and} \bibinfo{person}{Derek
  Dreyer}.} \bibinfo{year}{2018}\natexlab{b}.
\newblock \showarticletitle{{Iris from the ground up: {A} modular foundation
  for higher-order concurrent separation logic}}.
\newblock \bibinfo{journal}{\emph{J. Funct. Program.}}  \bibinfo{volume}{28}
  (\bibinfo{year}{2018}), \bibinfo{pages}{e20}.
\newblock
\urldef\tempurl%
\url{https://doi.org/10.1017/S0956796818000151}
\showDOI{\tempurl}


\bibitem[\protect\citeauthoryear{Kan, San{\'{a}}n, Lin, and Liu}{Kan
  et~al\mbox{.}}{2018}]%
        {DBLP:journals/corr/abs-1804-07608}
\bibfield{author}{\bibinfo{person}{Shuanglong Kan}, \bibinfo{person}{David
  San{\'{a}}n}, \bibinfo{person}{Shang{-}Wei Lin}, {and} \bibinfo{person}{Yang
  Liu}.} \bibinfo{year}{2018}\natexlab{}.
\newblock \showarticletitle{{K-Rust: An Executable Formal Semantics for Rust}}.
\newblock \bibinfo{journal}{\emph{CoRR}}  \bibinfo{volume}{abs/1804.07608}
  (\bibinfo{year}{2018}).
\newblock
\showeprint[arxiv]{1804.07608}
\urldef\tempurl%
\url{http://arxiv.org/abs/1804.07608}
\showURL{%
\tempurl}


\bibitem[\protect\citeauthoryear{Matsakis}{Matsakis}{2016}]%
        {MIR}
\bibfield{author}{\bibinfo{person}{Niko Matsakis}.}
  \bibinfo{year}{2016}\natexlab{}.
\newblock \bibinfo{title}{{Introducing {MIR}}}.
\newblock
  \bibinfo{howpublished}{\url{https://blog.rust-lang.org/2016/04/19/MIR.html}}.
\newblock


\bibitem[\protect\citeauthoryear{Matsakis}{Matsakis}{2017}]%
        {twophase-borrowing}
\bibfield{author}{\bibinfo{person}{Niko Matsakis}.}
  \bibinfo{year}{2017}\natexlab{}.
\newblock \bibinfo{title}{{Nested method calls via two-phase borrowing}}.
\newblock
  \bibinfo{howpublished}{\url{http://smallcultfollowing.com/babysteps/blog/2017/03/01/nested-method-calls-via-two-phase-borrowing/}}.
\newblock


\bibitem[\protect\citeauthoryear{M{\"{u}}ller, Schwerhoff, and
  Summers}{M{\"{u}}ller et~al\mbox{.}}{2016}]%
        {mcai/0001SS16}
\bibfield{author}{\bibinfo{person}{Peter M{\"{u}}ller}, \bibinfo{person}{Malte
  Schwerhoff}, {and} \bibinfo{person}{Alexander~J. Summers}.}
  \bibinfo{year}{2016}\natexlab{}.
\newblock \showarticletitle{{Viper: {A} Verification Infrastructure for
  Permission-Based Reasoning}}. In \bibinfo{booktitle}{\emph{Verification,
  Model Checking, and Abstract Interpretation - 17th International Conference,
  {VMCAI} 2016, St. Petersburg, FL, USA, January 17-19, 2016. Proceedings}}
  \emph{(\bibinfo{series}{Lecture Notes in Computer Science})},
  \bibfield{editor}{\bibinfo{person}{Barbara Jobstmann} {and}
  \bibinfo{person}{K.~Rustan~M. Leino}} (Eds.), Vol.~\bibinfo{volume}{9583}.
  \bibinfo{publisher}{Springer}, \bibinfo{pages}{41--62}.
\newblock
\showISBNx{978-3-662-49121-8}
\urldef\tempurl%
\url{https://doi.org/10.1007/978-3-662-49122-5_2}
\showDOI{\tempurl}


\bibitem[\protect\citeauthoryear{Olson}{Olson}{2016}]%
        {MIRI}
\bibfield{author}{\bibinfo{person}{Scott Olson}.}
  \bibinfo{year}{2016}\natexlab{}.
\newblock \bibinfo{title}{{M}iri}.
\newblock \bibinfo{howpublished}{\url{https://github.com/solson/miri}}.
\newblock


\bibitem[\protect\citeauthoryear{Podkopaev, Lahav, and Vafeiadis}{Podkopaev
  et~al\mbox{.}}{2019}]%
        {DBLP:journals/pacmpl/PodkopaevLV19}
\bibfield{author}{\bibinfo{person}{Anton Podkopaev}, \bibinfo{person}{Ori
  Lahav}, {and} \bibinfo{person}{Viktor Vafeiadis}.}
  \bibinfo{year}{2019}\natexlab{}.
\newblock \showarticletitle{{Bridging the gap between programming languages and
  hardware weak memory models}}.
\newblock \bibinfo{journal}{\emph{Proc. {ACM} Program. Lang.}}
  \bibinfo{volume}{3}, \bibinfo{number}{{POPL}} (\bibinfo{year}{2019}),
  \bibinfo{pages}{69:1--69:31}.
\newblock
\urldef\tempurl%
\url{https://doi.org/10.1145/3290382}
\showDOI{\tempurl}


\bibitem[\protect\citeauthoryear{Reed}{Reed}{2015}]%
        {Eric2015}
\bibfield{author}{\bibinfo{person}{Eric Reed}.}
  \bibinfo{year}{2015}\natexlab{}.
\newblock \bibinfo{booktitle}{\emph{{Patina: A Formalization of the Rust
  Programming Language}}}.
\newblock \bibinfo{type}{{T}echnical {R}eport}.
  \bibinfo{institution}{University of Washington}.
\newblock


\bibitem[\protect\citeauthoryear{Reynolds}{Reynolds}{2002}]%
        {DBLP:conf/lics/Reynolds02}
\bibfield{author}{\bibinfo{person}{John~C. Reynolds}.}
  \bibinfo{year}{2002}\natexlab{}.
\newblock \showarticletitle{{Separation Logic: {A} Logic for Shared Mutable
  Data Structures}}. In \bibinfo{booktitle}{\emph{17th {IEEE} Symposium on
  Logic in Computer Science {(LICS} 2002), 22-25 July 2002, Copenhagen,
  Denmark, Proceedings}}. \bibinfo{publisher}{{IEEE} Computer Society},
  \bibinfo{pages}{55--74}.
\newblock
\showISBNx{0-7695-1483-9}
\urldef\tempurl%
\url{https://doi.org/10.1109/LICS.2002.1029817}
\showDOI{\tempurl}


\bibitem[\protect\citeauthoryear{Ro{\c s}u and {\c S}erb{\u a}nu{\c t}{\u
  a}}{Ro{\c s}u and {\c S}erb{\u a}nu{\c t}{\u a}}{2010}]%
        {rosu-serbanuta-2010-jlap}
\bibfield{author}{\bibinfo{person}{Grigore Ro{\c s}u} {and}
  \bibinfo{person}{Traian~Florin {\c S}erb{\u a}nu{\c t}{\u a}}.}
  \bibinfo{year}{2010}\natexlab{}.
\newblock \showarticletitle{{An Overview of the {K} Semantic Framework}}.
\newblock \bibinfo{journal}{\emph{Journal of Logic and Algebraic Programming}}
  \bibinfo{volume}{79}, \bibinfo{number}{6} (\bibinfo{year}{2010}),
  \bibinfo{pages}{397--434}.
\newblock
\urldef\tempurl%
\url{https://doi.org/10.1016/j.jlap.2010.03.012}
\showDOI{\tempurl}


\bibitem[\protect\citeauthoryear{Rust-Benchmark}{Rust-Benchmark}{2020}]%
        {benchmark}
\bibfield{author}{\bibinfo{person}{Rust-Benchmark}.}
  \bibinfo{year}{2020}\natexlab{}.
\newblock \bibinfo{title}{Rust Benchmark}.
\newblock
  \bibinfo{howpublished}{\url{https://github.com/rust-lang/rust/tree/master/src/test}}.
\newblock


\bibitem[\protect\citeauthoryear{Rust-Team}{Rust-Team}{2016}]%
        {rusthome}
\bibfield{author}{\bibinfo{person}{Rust-Team}.}
  \bibinfo{year}{2016}\natexlab{}.
\newblock \bibinfo{title}{The Rust Language Homepage}.
\newblock \bibinfo{howpublished}{\url{https://www.rust-lang.org/en-US/}}.
\newblock


\bibitem[\protect\citeauthoryear{Rust-Team}{Rust-Team}{2018}]%
        {NLL2018}
\bibfield{author}{\bibinfo{person}{Rust-Team}.}
  \bibinfo{year}{2018}\natexlab{}.
\newblock \bibinfo{title}{{Non-lexical lifetimes}}.
\newblock
  \bibinfo{howpublished}{\url{https://doc.rust-lang.org/edition-guide/rust-2018/ownership-and-lifetimes/non-lexical-lifetimes.html}}.
\newblock


\bibitem[\protect\citeauthoryear{{Rust-Team}}{{Rust-Team}}{2018}]%
        {rustbook}
\bibfield{author}{\bibinfo{person}{{Rust-Team}}.}
  \bibinfo{year}{2018}\natexlab{}.
\newblock \bibinfo{booktitle}{\emph{{The Rust Programming Language}}}.
\newblock Mozilla Research.
\newblock
\newblock
\shownote{https://doc.rust-lang.org/book/foreword.html.}


\bibitem[\protect\citeauthoryear{Stork, Naden, Sunshine, Mohr, Fonseca,
  Marques, and Aldrich}{Stork et~al\mbox{.}}{2014}]%
        {DBLP:conf/pldi/StorkNSMFMA14}
\bibfield{author}{\bibinfo{person}{Sven Stork}, \bibinfo{person}{Karl Naden},
  \bibinfo{person}{Joshua Sunshine}, \bibinfo{person}{Manuel Mohr},
  \bibinfo{person}{Alcides Fonseca}, \bibinfo{person}{Paulo Marques}, {and}
  \bibinfo{person}{Jonathan Aldrich}.} \bibinfo{year}{2014}\natexlab{}.
\newblock \showarticletitle{{\AE}minium: {a permission based
  concurrent-by-default programming language approach}}. In
  \bibinfo{booktitle}{\emph{{ACM} {SIGPLAN} Conference on Programming Language
  Design and Implementation, {PLDI} '14, Edinburgh, United Kingdom - June 09 -
  11, 2014}}, \bibfield{editor}{\bibinfo{person}{Michael F.~P. O'Boyle} {and}
  \bibinfo{person}{Keshav Pingali}} (Eds.). \bibinfo{publisher}{{ACM}},
  \bibinfo{pages}{26}.
\newblock
\urldef\tempurl%
\url{https://doi.org/10.1145/2594291.2594344}
\showDOI{\tempurl}


\bibitem[\protect\citeauthoryear{Tofte and Talpin}{Tofte and Talpin}{1994}]%
        {DBLP:conf/popl/TofteT94}
\bibfield{author}{\bibinfo{person}{Mads Tofte} {and}
  \bibinfo{person}{Jean{-}Pierre Talpin}.} \bibinfo{year}{1994}\natexlab{}.
\newblock \showarticletitle{{Implementation of the Typed Call-by-Value
  lambda-Calculus using a Stack of Regions}}. In
  \bibinfo{booktitle}{\emph{Conference Record of POPL'94: 21st {ACM}
  {SIGPLAN-SIGACT} Symposium on Principles of Programming Languages, Portland,
  Oregon, USA, January 17-21, 1994}},
  \bibfield{editor}{\bibinfo{person}{Hans{-}Juergen Boehm},
  \bibinfo{person}{Bernard Lang}, {and} \bibinfo{person}{Daniel~M. Yellin}}
  (Eds.). \bibinfo{publisher}{{ACM} Press}, \bibinfo{pages}{188--201}.
\newblock
\urldef\tempurl%
\url{https://doi.org/10.1145/174675.177855}
\showDOI{\tempurl}


\bibitem[\protect\citeauthoryear{Toman, Pernsteiner, and Torlak}{Toman
  et~al\mbox{.}}{2015}]%
        {TomanPT15}
\bibfield{author}{\bibinfo{person}{John Toman}, \bibinfo{person}{Stuart
  Pernsteiner}, {and} \bibinfo{person}{Emina Torlak}.}
  \bibinfo{year}{2015}\natexlab{}.
\newblock \showarticletitle{{Crust: {A} Bounded Verifier for Rust {(N)}}}. In
  \bibinfo{booktitle}{\emph{30th {IEEE/ACM} International Conference on
  Automated Software Engineering, {ASE} 2015, Lincoln, NE, USA, November 9-13,
  2015}}, \bibfield{editor}{\bibinfo{person}{Myra~B. Cohen},
  \bibinfo{person}{Lars Grunske}, {and} \bibinfo{person}{Michael Whalen}}
  (Eds.). \bibinfo{publisher}{{IEEE} Computer Society},
  \bibinfo{pages}{75--80}.
\newblock
\showISBNx{978-1-5090-0025-8}
\urldef\tempurl%
\url{https://doi.org/10.1109/ASE.2015.77}
\showDOI{\tempurl}


\bibitem[\protect\citeauthoryear{Tov and Pucella}{Tov and Pucella}{2011}]%
        {DBLP:conf/popl/TovP111}
\bibfield{author}{\bibinfo{person}{Jesse~A. Tov} {and}
  \bibinfo{person}{Riccardo Pucella}.} \bibinfo{year}{2011}\natexlab{}.
\newblock \showarticletitle{Practical affine types}. In
  \bibinfo{booktitle}{\emph{Proceedings of the 38th {ACM} {SIGPLAN-SIGACT}
  Symposium on Principles of Programming Languages, {POPL} 2011, Austin, TX,
  USA, January 26-28, 2011}}, \bibfield{editor}{\bibinfo{person}{Thomas Ball}
  {and} \bibinfo{person}{Mooly Sagiv}} (Eds.). \bibinfo{publisher}{{ACM}},
  \bibinfo{pages}{447--458}.
\newblock
\urldef\tempurl%
\url{https://doi.org/10.1145/1926385.1926436}
\showDOI{\tempurl}


\bibitem[\protect\citeauthoryear{Wang, Song, Zhang, Zhu, and Zhang}{Wang
  et~al\mbox{.}}{2018}]%
        {corr/abs-1804-10806}
\bibfield{author}{\bibinfo{person}{Feng Wang}, \bibinfo{person}{Fu Song},
  \bibinfo{person}{Min Zhang}, \bibinfo{person}{Xiaoran Zhu}, {and}
  \bibinfo{person}{Jun Zhang}.} \bibinfo{year}{2018}\natexlab{}.
\newblock \showarticletitle{{KRust: {A} Formal Executable Semantics of Rust}}.
\newblock \bibinfo{journal}{\emph{CoRR}}  \bibinfo{volume}{abs/1804.10806}
  (\bibinfo{year}{2018}).
\newblock
\showeprint[arxiv]{1804.10806}
\urldef\tempurl%
\url{http://arxiv.org/abs/1804.10806}
\showURL{%
\tempurl}


\bibitem[\protect\citeauthoryear{Weiss, Patterson, Matsakis, and Ahmed}{Weiss
  et~al\mbox{.}}{2019}]%
        {DBLP:journals/corr/abs-1903-00982}
\bibfield{author}{\bibinfo{person}{Aaron Weiss}, \bibinfo{person}{Daniel
  Patterson}, \bibinfo{person}{Nicholas~D. Matsakis}, {and}
  \bibinfo{person}{Amal Ahmed}.} \bibinfo{year}{2019}\natexlab{}.
\newblock \showarticletitle{{Oxide: The Essence of Rust}}.
\newblock \bibinfo{journal}{\emph{CoRR}}  \bibinfo{volume}{abs/1903.00982}
  (\bibinfo{year}{2019}).
\newblock
\showeprint[arxiv]{1903.00982}
\urldef\tempurl%
\url{http://arxiv.org/abs/1903.00982}
\showURL{%
\tempurl}


\end{thebibliography}

\appendix

\section{Proof of Lemma \ref{lemma-acyclic}}
\label{sec-acyclic}
\begin{proof}
	Assume there is a cycle in a well-formed OBS graph,
	there must exist $a\rightarrow_b a'$ and $a'\rightarrow_b a_1\ldots a_2\rightarrow_b a$.
	We can infer that $\mathcal{F}(a\rightarrow_b a')\subset\mathcal{F}(a_2\rightarrow_b a)$
	and $\mathcal{F}(a_2\rightarrow_b a)\subset\mathcal{F}(a\rightarrow_b a')$ by the condition (2) of Definition \ref{def-well-formed-obs}, which cannot be true, 
	
\end{proof}
\section{Proof of Theorem \ref{theorem-exclusive}}
\label{app-exclusive}
\begin{proof}
	This theorem specifies that exclusive mutation guarantee are maintained in a well-formed OBS graph with respect to the permission functions.

	If the owner and all references are shared aliases then the theorem is trivially proved.
	If there is a mutable alias enabled at the  {timestamp $t$} then
	it is proved by two cases.
	
	\emph{Case 1:} Assume the owner is the mutable alias. 
	We can infer that there is no reference that has the writing or reading permission to  the memory location at {$t$}.
	Otherwise the reference should disable the owner according to 
	the definition of the permission functions
	and the unique owner invariant of well-formed OBS graphs.
	
	\emph{Case 2:} Assume there is a mutable reference $a$ to the memory block enabled at {$t$}, and the owner of the memory block is $x$, we have that
	there exists a link $a\rightarrow_m a_1 \ldots  x\rightarrow_o B $. The owner and all other references in the link should be disabled
	by  $a$ and thus their writing permissions are not enabled.
	
	Now assume there is another reference $a'\neq a$ enabled at {$t$}. 
	It should not be in the link from $a$ to $x$.
	We have that there exists another  link 
	$a'\rightarrow_m a_1' \ldots x{\rightarrow_o} B$.
	Then there must exists $a_2,a_2',a_3$ such that $a\rightarrow_m a_1 \ldots a_2\rightarrow_b a_3\ldots x\rightarrow_{o} B$ and
	$a'\rightarrow_m a_1' \ldots a_2'\rightarrow_{b} a_3\ldots x\rightarrow_o B$
	(See Fig. \ref{fig-theorem-unfiedmodel}).
	According to Invariant (3) of the definition for well-formed OBS graphs (Definition \ref{def-well-formed-obs}), the lifetimes
	of $a_2\rightarrow_b a_3$ and $a_2'\rightarrow_{b} a_3$ should not intersect.
	According to Invariant (2) of Definition \ref{def-well-formed-obs}, 
	the lifetime of $a\rightarrow_m a_1$ should be within the lifetime of $a_2\rightarrow_b a_3$ and
	the lifetime of $a'\rightarrow_m a_1'$ should be within the lifetime of $a_2'\rightarrow_b a_3$.
	Thus
	the lifetimes of $a\rightarrow_m a_1$ and $a' \rightarrow_m a_1'$ cannot intersect.
	Therefore the lifetime of $a'\rightarrow_m a_1'$ does not contain the current timestamp  {$t$} and $a'$ is not permitted to write.
\end{proof}

\begin{figure}
	\centering
	\includegraphics[scale=0.4]{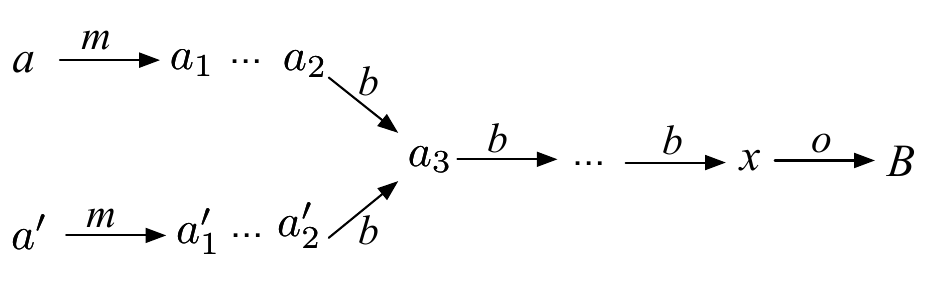}
	\caption{The proof idea of Theorem \ref{theorem-unifimodel}}
	\label{fig-theorem-unfiedmodel}
\end{figure}

\section{Proof of Theorem \ref{theorem-refinement}}
\label{append-thm-refinement}

The first claim that $G$ is a well-formed OBS graph is proved by 
Lemma \ref{lemma-memlayout} and \ref{lemma-lft}.
Lemma \ref{lemma-memlayout} proves that each memory configuration in a safe sequence is a well-formed configuration and 
Lemma \ref{lemma-lft} proves the lifetime inclusion variant for well-formed configurations.
The second claim is proved by Lemma \ref{lemma-mut} and \ref{lemma-shr}.

\begin{lemma}\label{lemma-memlayout}
Let ${\pi}= (mem_0,\cd{ts}_0),op_1,(mem_1,\cd{ts}_1),\ldots, (mem_{n-1}, \cd{ts}_{n-1}),op_n,(mem_{n}, \cd{ts}_{n})$ be a safe sequence.
Then all $mem_i,0\leq i \leq n$, are well-formed. 
\end{lemma}

\begin{proof}
	Since $m_0$ is an empty configuration, it is trivially well-formed.
	The allocation operation does not change the well-formedness of a memory layout.
	The free operation also does not change the well-formedness of a memory {configuration}, since it only removes a block from the heap.
	For reading and writing operations, the semantics rules always ensure the resulting 
	memory {configurations} are well-formed.
\end{proof}

\begin{lemma}\label{lemma-disable}
	Let ${\pi}= (mem_0,\cd{ts}_0),op_1,(mem_1,\cd{ts}_1),\ldots, (mem_{n-1}, \cd{ts}_{n-1}),op_n,(mem_{n}, \cd{ts}_{n})$ be a safe sequence.
	Let $mem_i,0\leq i \leq n$ be a memory configuration..
	We have:
	\begin{enumerate}
		\item For any $\cd{a}\rightarrow_s\cd{a}'\in mem_i$,  there is no writing
		by $\cd{a}'$ during the lifetime of $\cd{a}$ in $mem_i$.
		\item For any $\cd{a}\rightarrow_m\cd{a}'\in mem_i$,  there is no reading or writing
		by $\cd{a}'$ during the lifetime of $\cd{a}$ in $mem_i$.
	\end{enumerate}
	
\end{lemma}

\begin{proof}
	We first prove the claim (1).
	  Assume  $\cd{a}$'s lifetime is $\cd{ts}_b\sim \cd{ts}_e$ in $mem_i$ (i.e., $\mathcal{L}(\cd{a},mem_i)=\cd{ts}_b\sim \cd{ts}_e$) and 
	\cd{ts} is a timestamp at which $\cd{a}'$ is used to write and $\cd{ts}\in \cd{ts}_b\sim \cd{ts}_e$.

	The writing by $\cd{a}'$ at \cd{ts} should not be the latest writing in $mem_i$, that is, if $mem_i=(S,H,\mathcal{P},ms)$  
	and $\mathcal{P}(\cd{a}')=(\_,\cd{ts}')$ then $\cd{ts}\neq \cd{ts}'$, otherwise it should be disabled according to Definition \ref{def-wellformlayouts}.
	Therefore we have $\cd{ts}' >\cd{ts}_e$.
	
	{Without loss of generality, we assume there is no writing by $\cd{a}'$ during $\cd{ts}\sim\cd{ts}_e$.}
	Then we have that the memory configuration $mem_k$ in $(mem_k,\cd{ts}_k)$ with $\cd{ts}_k=\cd{ts}_e$
	is not a well-formed memory configuration as 
	the writing by $\cd{a}'$ at $\cd{ts}$ is the lasting writing by $\cd{a}'$ in $mem_k$ but
	the lifetime of \cd{a} in $mem_k$ is $\cd{ts}_b\sim \cd{ts}_e$.
	The writing should be disabled according to Definition \ref{def-wellformlayouts}.
	This contradicts to Lemma \ref{lemma-disable}.
\end{proof}

\begin{lemma}\label{lemma-lft}
	Let ${\pi}= (mem_0,\cd{ts}_0),op_1,(mem_1,\cd{ts}_1),\ldots, (mem_{n-1}, \cd{ts}_{n-1}),op_n,(mem_{n}, \cd{ts}_{n})$ be a safe sequence.
	Let $mem_i,0\leq i \leq n$ be a memory configuration and $B$ be a block with the location $\cd{b}$.
	Let the OBS graph of $B$ in $mem_i$ be $G=(V,B,E,\mathcal{F})$. We have that $G$ is well-formed. 
\end{lemma}

\begin{proof}
	From Lemma \ref{lemma-memlayout}, $mem_i$ is well-formed, which
	{ensures} that $G$ satisfies the \emph{unique owner invariant} and
	\emph{no intersection invariant}.
	We still need to prove the \emph{lifetime inclusion invariant}.
	 
	 For references, according the definition of $\mathcal{L}(\cd{a},mem)$, we know that 
	the lifetime of \cd{a} will always include the
	 lifetime of $\cd{a}'$, where $\cd{a}'\rightarrow^b_* \cd{a}\in mem$.

	We need to consider the owner. If 
	an owner's writing permission is always disabled by its references then the ownership cannot be moved.
	In this case, its lifetime always includes it's references' lifetimes.
	From Lemma \ref{lemma-disable}, we know that the owner is always disabled.
%
%
%
%
%
%
%
\end{proof}


\begin{lemma}\label{lemma-mut}
	Let ${\pi}= (mem_0,\cd{ts}_0),op_1,(mem_1,\cd{ts}_1),\ldots, (mem_{n-1}, \cd{ts}_{n-1}),op_n,(mem_{n}, \cd{ts}_{n})$ be a safe sequence.
	Let $G=(V,B,E,\mathcal{F})$ be the OBS graph of $B$ in $mem_i,0\leq i \leq n$, we have
	for any two nodes $a,a'\in V\cup\{B\}$, if a reading operation $\cd{read}(p)$, where $\cd{alias}(p)=a$,
	is carried out at the timestamp  $\cd{ts}\in \mathcal{F}(a\rightarrow_* a')$ then
	$R_G(a,\cd{ts})=true$.
\end{lemma}

\begin{proof}
	
	The {key} to prove the lemma is to ensure that during the lifetime of $a$, the reading by $a$ at $\cd{ts}$ is not disabled. Actually it is proved by Lemma \ref{lemma-disable}.
	
%
%
%
%
 
\end{proof}

\begin{lemma}\label{lemma-shr}
	Let ${\pi}= (mem_0,\cd{ts}_0),op_1,(mem_1,\cd{ts}_1),\ldots, (mem_{n-1}, \cd{ts}_{n-1}),op_n,(mem_{n}, \cd{ts}_{n})$ be a safe sequence.
Let $G=(V,B,E,\mathcal{F})$ be the OBS graph of $B$ in $mem_i,0\leq i \leq n$, we have
for any two nodes $a,a'\in V\cup\{B\}$, if a writing operation $\cd{write}(p,v)$, where $\cd{alias}(p)=a$,
is carried out at the timestamp  $\cd{ts}\in \mathcal{F}(a\rightarrow_* a')$ then
$W_G(a,\cd{ts})=true$.
\end{lemma}
\begin{proof}
	Proof is similar to Lemma \ref{lemma-mut}.
\end{proof}

Theorem \ref{theorem-refinement} is proved by Lemma \ref{lemma-memlayout}, \ref{lemma-lft},
\ref{lemma-mut}, and \ref{lemma-shr}.

\end{document}